\documentclass[aps,preprint,12pt,eqsecnum,nofootinbib,amsmath,amssymb]{revtex4}

\newcommand\mydates{19 June 2007}

\usepackage{graphicx} 
\usepackage{bm}       
\usepackage{latexsym} 

\newcommand{\footnoteskip}{\baselineskip 12pt plus 1pt minus 1pt}
\newcommand{\captionskip}{\footnotesize \baselineskip 12pt plus 1pt minus 1pt}
\newcommand{\tableofcontentsskip}{\baselineskip 14pt plus 1pt minus 1pt}
\newcommand{\affiliationskip}{\baselineskip 15pt plus 1pt minus 1pt}
\newcommand{\titleskip}{\baselineskip 18pt plus 1pt minus 1pt}
\newcommand{\abstractskip}{\baselineskip 13pt plus 1pt minus 1pt}
\newcommand{\bodyskip}{\baselineskip 18pt plus 1pt minus 1pt}
\newcommand{\bibliographyskip}{\baselineskip16pt plus 1pt minus 1pt}
\newcommand{\enumerateskip}{\baselineskip 14pt plus 1pt minus 1pt}

\newcommand{\smGT}{{\scriptscriptstyle >}}
\newcommand{\smLT}{{\scriptscriptstyle <}}

\newcommand{\smBPS}{{\rm\scriptscriptstyle BPS}}

\newcommand{\smD}{{\rm\scriptscriptstyle D}}
\newcommand{\smI}{{\rm\scriptscriptstyle I}}
\newcommand{\smB}{{\rm\scriptscriptstyle B}}

\newcommand{\smH}{{\rm\scriptscriptstyle H}}
\newcommand{\smCoul}{{\rm\scriptscriptstyle coul}}
\newcommand{\smPerp}{{\rm\scriptscriptstyle \perp}}

\begin{document}

\preprint{LA-UR-06-6738}

\title{\titleskip
  BPS Explained I: Temperature Relaxation in a Plasma \\[5pt]
  or\\[5pt]
  How to Find the Coulomb Logarithm Exactly
}

\author{Robert~L. Singleton Jr.}

\vskip0.2cm 
\affiliation{\affiliationskip
     Los Alamos National Laboratory\\
     Los Alamos, New Mexico 87545, USA
}
\date{\mydates}

\begin{abstract}
\abstractskip
\vskip0.3cm 

\noindent
  This is the first of two lectures on the technique of
  dimensional continuation employed by Brown, Preston, and Singleton
  (BPS) to calculate such quantities as the charged particle stopping
  power and the temperature equilibration rate in a plasma. In this
  exposition we will examine~some of the more basic points of
  dimensional continuation, with an emphasis on the Coulomb logarithm
  for electron-ion temperature equilibration. Dimensional
  continuation, or dimensional regularization as it is more properly
  known in quantum field theory, was originally developed as part of
  the renormalization procedure for the theories of the electroweak
  and other fundamental interactions in particle physics.  Dimensional
  continuation is so general, in fact, that any {\em theory} can be
  unambiguously lifted to dimensions beyond three, and therefore the
  technique is powerful enough to apply in many other settings. The
  technique, however, is not well known outside the field theory and
  particle physics communities. This exposition will therefore be
  self-contained, intended for those who are not specialists in
  quantum field theory, and I will either derive or motivate any
  requisite field theory results or concepts. Of particular relevance
  is the analogy between the Coulomb logarithm as calculated by Lyman
  Spitzer on the one hand, and the Lamb shift as calculated by Hans
  Bethe on the other. While dimensional continuation is a well
  developed and a thoroughly tested method for regularizing any
  quantum field theory, BPS employs the method in a novel way that
  provides the leading and subleading behavior for processes that
  involve competing disparate energy or length scales. In particular,
  BPS calculated the temperature equilibration rate to leading and
  next-to-leading order in the plasma number density for any two
  species in a plasma that are in thermal equilibrium with themselves,
  but not necessarily with each other. No restriction is made on the
  charge, mass, or the temperature of the plasma species. It is,
  however, assumed that the plasma is not strongly coupled in the
  sense that the dimensionless plasma coupling parameter $g= e^2
  \kappa_\smD / 4 \pi T $ is small, where $\kappa_\smD$ is the Debye
  wave number of the plasma. To leading and next-to-leading order in
  this coupling, the temperature equilibration rate is of the generic
  form $\Gamma = A\, g^2 \, \ln\{C g\}$. The precise numerical
  coefficient $A$ in front of the logarithm has been known for some
  time, while BPS have recently computed the constant $C$ under the
  logarithm.  It should be emphasized that the BPS result is not a
  model, but rather it is an exact calculation of the leading terms in
  a well-defined perturbation theory.  This exact result differs from
  approximations and models given in the literature.
\end{abstract}

\maketitle

\pagebreak
\tableofcontentsskip
\tableofcontents

\thispagestyle{empty}

\pagebreak
\bodyskip
\section{Introduction and Context}

\setcounter{page}{1}

This is the first of two lectures on a new technique for
calculating the temperature equilibration rate between electrons and
ions in a weakly to moderately coupled fully-ionized plasma, exact to
leading and next-to-leading order in the plasma number density.  This
calculation was first performed in Section~12 of Ref.~\cite{bps}, a
work whose primary focus was the charged particle stopping power in a
plasma.\footnote{\footnoteskip 
  For a short cursory version of this work, see Ref.~\cite{bpsshort}.  
} 
This paper assumed familiarity with a number of field theory concepts,
and Section~12 relied heavily on the charged particle stopping power
results derived in previous sections of that work. In contrast, these
lectures will be self-contained. I will either derive or motivate the
requisite field theory background for a complete reading of
Ref.~\cite{bps}, with an emphasis on some of the subtleties of the
calculational techniques and the concepts behind them (this lecture).
More to the point, since the rate calculation stands on its own, it
should be presented on its own (the following lecture). Indeed, since
the calculation of the rate is somewhat less involved than that of the
stopping power, it more clearly illustrates the tools and concepts
imported from field theory.

In addition to clarifying the method of dimensional continuation, this
lecture will place Ref.~\cite{bps} in the context of more familiar and
traditional approaches to the rate problem. In particular, I will show
that dimensional continuation can be viewed as a systematic
implementation of the approach based on convergent kinetic equations.
Finally, in in the next lecture, I will go on to derive the main
result from Section~12 of Ref.~\cite{bps}, the rate coefficient
(\ref{bpsrate}) of this lecture.  By working in the Born
approximation, and adopting the methods of Ref.~\cite{bs}, I will
derive this result in a much simpler manner than originally presented
in Ref.~\cite{bps}.  While Lectures~I~and~II are self-contained, they
are complementary and should be read as a unit.\footnote{\footnoteskip
  In other words, Lecture~I (this lecture) consists of the basic
  theory and techniques behind dimensional continuation, while
  Lecture~II (next lecture) will be a specific calculation in the
  extreme quantum limit:~only by performing a calculation can one
  understand the underlying ideas of the calculation.  Ideally, I
  would then like to continue these first two lectures with three
  additional ones. In Lecture~III, I would present the full
  calculation of the temperature relaxation rate performed in
  Ref.~\cite{bps}, valid beyond the Born approximation.  The full
  calculation is accurate for a weakly to moderately coupled plasma in
  both the classical and quantum limits, and any regime in-between,
  regardless of the mass and temperature difference of the plasma
  species.  I would then present the details of the full quantum
  corrections in Lecture~IV. Finally, in Lecture~V, I would come full
  circle, simplifying the general equilibration rate calculated in
  Lectures~III and IV to obtain the extreme quantum limit of Lecture
  II, equation (\ref{bpsrate}) of this lecture. This would provide two
  independent calculations of the rate (\ref{bpsrate}), but alas, time
  does not permit these last three supplements.
}

The strategy employed by Ref.~\cite{bps}, hereafter referred to
as~BPS, consists of two steps. First, we will find a dimensionless
parameter in which to perform a {\em controlled} perturbative
expansion of the rate, expanding to leading and next-to-leading order
in this parameter. Second, we will deploy a technique from quantum
field theory that will allow us to calculate 
the coefficients of these leading and
subleading terms {\em exactly}.  The exact leading order term is not
very difficult to find, and has been known since the classic work of
Spitzer. The next-to-leading order term, on the other hand, was not
known exactly until the recent BPS calculation. The third-order term
provides an estimate of the error of the calculation. When the plasma
is weakly to moderately coupled, the error will be small and the rate
will be approximated quite accurately by the first two terms of this
expansion.

To calculate the expansion coefficients, we will exploit a field theory 
technique known as dimensional regularization (or dimensional continuation, 
as I will call it here). This application of dimensional continuation is 
quite different from its intended purpose in the renormalization procedure 
of quantum field theory. Dimensional continuation was originally developed 
as an elegant regularization scheme in which the fundamental symmetries of 
a field theory could be maintained while still rendering finite the otherwise 
infinite integrals that arise when calculating Feynman diagrams. I will show 
how this technique can be used in a novel fashion to extract next-to-leading 
order physics that has, until now, remained inaccessible. In other words, 
I will show how dimensional continuation provides an exact result for the 
corresponding Coulomb logarithm of the process in question. I will also take 
the opportunity to correct a small algebra mistake for the electron-ion 
equilibration rate presented in Section~12 of Ref.~\cite{bps}.

\subsection{The Problem}

The general formalism starts with a 
plasma composed of multiple species labeled by an index $b$, the 
various species being delineated by of a common electric charge 
$e_b$ and a common mass $m_b$.\footnote{\footnoteskip
 The final constraint of BPS is that the ionization state of each
 component species does not change.  In other words, the charges can
 be expressed as $e_b=Z_b\, e$ with $Z_b$ fixed.  While this
 simplifying assumption has its limitations, it facilitates the
 analytic calculation of the stopping power and temperature
 equilibration rate.  However, one can take the $Z_b$ to be fractional
 to mock-up ionization in a simple manner. Or better yet, it should be
 possible to combine the BPS results with models of the ionization
 effects. For a hot low-$Z$ plasmas, such as a deuterium-tritium
 plasma during ICF ignition, the ions are likely to be fully ionized
 in any event, so this is not a serious restriction for clean
 thermonuclear burn with low-$Z$ impurities.
}
Each species is assumed to be in thermal equilibrium with itself at
temperature $T_b$ with a spatially uniform number density $n_b$. I
will drop the subscript on the charge of the electron and
write $e_\text{electron}=-e$ (with $e>0$), although the electron mass
will be denoted by $m_e$, and the number density and temperature of
the electron plasma component by $n_e$ and $T_e$ respectively. I will
use a lower case subscript $i$ to denote a single ion species of
charge $e_i = Z_i \, e$, mass $m_i$, number density $n_i$, and
temperature $T_i$.  I will employ a capital-I subscript to denote
properties that correspond to the collective set of ions, such as
the total ion number density $n_\smI={\sum}_i n_i$ or a common ion
temperature~$T_\smI$.  When equilibrium distributions are required in
calculations, they are assumed to be Maxwell-Boltzmann, although a
generalization to Fermi-Dirac statistics can be accomplished with
more effort~\cite{bs}.  For problems involving hot thermonuclear
burn, however, the fugacity is small and Maxwell-Boltzmann statistics
is an accurate approximation.

Let $d{\cal E}_{ab}/dt$ denote rate at which the energy density of
plasma species $a$ changes because of its Coulomb interactions with
another species $b$ (the rate {\em from} the $a$-species {\em to} 
the $b$-species). This rate is proportional to the temperature
difference, and can be expressed by
\begin{eqnarray}
  \frac{d{\cal E}_{ab}}{dt}
  =   -\, {\cal C}_{ab}\left(T_a-T_b\right) \ .
\label{dedtei}
\end{eqnarray}
The sign convention in (\ref{dedtei}) implies that when the rate
coefficients ${\cal C}_{ab}$ are positive, then energy will flow from
the hotter species to the cooler species, as it must. Section~12 of
BPS used dimensional continuation to calculate the general rate
coefficients ${\cal C}_{ab}$ in a weakly coupled, but otherwise
arbitrary, plasma. For simplicity, we will not perform the general
calculation until Lecture~III. In this and the following lecture, we
will concentrate on the energy exchange between electrons and ions
only. Since the electron mass $m_e$ is so much smaller than a typical
ion mass~$m_\smI$, the electrons will come into equilibrium first with
temperature $T_e$ on some time scale~$\tau_e$. The energy transfer
rate among ions is a factor $\sqrt{m_e/m_\smI}$ slower than the
corresponding rate for electrons, and therefore the ions will
equilibrate to a common temperature $T_\smI$ in a time $\tau_\smI \sim
\tau_e\,\sqrt{m_\smI/m_e}$\,.  Finally, as the electrons and ions
exchange energy through Coulomb interactions, these systems too will
equilibrate on a time scale \hbox{$\tau_{e\smI} \sim \tau_\smI\,
\sqrt{m_\smI/m_e} \sim \tau_e\,(m_\smI/m_e)$}. Consequently, one finds
a hierarchy of time scales \hbox{$\tau_e \ll \tau_\smI \ll \tau_{e
\smI}$}, and it indeed makes sense to consider the electron and ion
systems as having distinct temperatures $T_e$ and $T_\smI$, with
subsequent equilibration between them. Taking $a=e$ and $b=i$ in
(\ref{dedtei}), and since the ions have a common temperature
$T_i=T_\smI$, the rate equation of interest is obtained by summing
over the ion components of (\ref{dedtei}) to give
\begin{eqnarray}
  \frac{d{\cal E}_{e\smI}}{dt}
  =   -\, {\cal C}_{e\smI}\left(T_e-T_\smI\right) \ ,
\label{dedteI}
\end{eqnarray}
where ${\cal C}_{e\smI}=\sum_i {\cal C}_{ei}$ and $d{\cal
E}_{e\smI}/dt = {\sum}_i\, d{\cal E}_{e i}/dt$.

The coefficient ${\cal C}_{e\smI}$ is the quantity we wish to
calculate in this and the next lecture. This coefficient contains the
energy-exchange physics between electrons and ions resulting from 
mutual Coulomb
interactions, including possible collective effects and large-angle
collisions. General expressions for the individual ${\cal C}_{e i}$
were calculated in Section~12 of BPS. They are somewhat complicated
and involve various one-dimensional integrals that can only be
performed numerically. However, the collective rate coefficient ${\cal
C}_{e\smI}$ simplifies considerably when the mild 
restriction \hbox{ $m_e/T_e \ll m_\smI/T_\smI$}\, is imposed (a
sum-rule is employed in the approximation, and the simplification
occurs only for ${\cal C}_{e\smI} = \sum_i {\cal C}_{e i}$ and not for
the individual ${\cal C}_{e i}$).  If the high temperature limit is
further imposed, then the rate can be written in a quite simple
analytic form.
\begin{eqnarray}
  && \text{For } m_e/T_e \ll m_\smI/T_\smI  
  ~~~ \text{and} ~~~ T_{e,\smI} \gg \epsilon_\smH \,:
\nonumber
\\[5pt]
  &&{\cal C}_{e\smI}
  = 
  \frac{\kappa_e^2}{2\pi}\, \omega_\smI^2\,
  \sqrt{\frac{m_e}{2\pi\, T_e}}\, \ln\Lambda_\smBPS \ ,
  ~~~\text{with}~~~
  \ln\Lambda_\smBPS
  =
  \frac{1}{2}\left[\ln\!\left\{\frac{8 T_e^2}{\hbar^2 \omega_e^2}
  \right\} - \gamma - 1 \right] \ ,
\label{bpsrate}
\end{eqnarray}
\vskip0.5cm 
\noindent 
where $\gamma=0.57721 \cdots$ is the Euler constant, $\kappa_e$ and
$\omega_e$ are the electron Debye wave number and plasma frequency,
and \hbox{$\omega_\smI^2 =\sum_i \omega_i^2$} is sum of the squares of
the ion plasma frequencies. Since the small binding energy $\epsilon_\smH =
13.6\, {\rm eV}$ of the hydrogen atom sets the temperature scale, and
since the condition $m_e/T_e \ll m_\smI/T_\smI$ is not very
restrictive, the rate coefficient (\ref{bpsrate}) is applicable in
almost all circumstances of interest. We shall devote the next lecture
to deriving this expression. For now, note that
equation~(\ref{bpsrate}) corresponds to Eqs.~(3.61)~and~(12.12) of
Ref.~\cite{bps}, where I have taken this opportunity to correct a
small transcription error: when passing from Eq.~(12.43) to
Eq.~(12.44) in Ref.~\cite{bps}, a factor of 1/2 was dropped. Restoring
this factor of 1/2 changes the additive constant outside the logarithm
from the $-\gamma-2$ that appears in Eq.~(12.12) of Ref.~\cite{bps} to
the constant $-\gamma-1$ in~(\ref{bpsrate}) above. 

For reasons to be discussed shortly, rationalized units are
preferred for dimensional continuation, and I will employ this choice
in all that follows. Nonetheless, expression (\ref{bpsrate}) is
written in a manner that does not depend upon this choice: the Debye
wave number $\kappa_e$, and the plasma frequencies $\omega_e$ and
$\omega_i$ can be calculated in your favorite units. For example, in
Gaussian units where the electric potential takes the form
$V=e^2/r$, the Debye wave number and the plasma frequency of species
$b$ are given by \hbox{$\kappa_b^2 = 4\pi\, e_b^2\, n_b/T_b$} and
\hbox{$\omega_b^2 = 4\pi\, e_b^2\, n_b/m_b$}.  In the rationalized
units employed here, the electric potential is given by $V=e^2/4\pi
r$, and we have
\begin{eqnarray}
  \kappa_b^2 &=& \frac{e_b^2\, n_b}{T_b}
\label{kbdef}
\\[5pt]
  \omega_b^2 &=& \frac{e_b^2\, n_b}{m_b} \ .
\label{wbdef}
\end{eqnarray}
The square of the total Debye wave number is $\kappa_\smD^2 =\sum_b
\kappa_b^2$, and the total Debye wave length is $\lambda_\smD=
\kappa_\smD^{-1}$.

\pagebreak
\subsection{The Problem with the Problem}
\label{sec:problemwithproblem}

Let us now consider an arbitrary plasma component of mass $m$, which I 
will otherwise leave unspecified, and let $f({\bf v},t)$ denote the Boltzmann 
distribution for this species. Then the average (kinetic) energy density 
of this component is
\begin{eqnarray}
  {\cal E}
  = 
  \int d^3 v \,\frac{1}{2}\, m v^2\, f({\bf v},t) \ .
\label{Ebkin}
\end{eqnarray}
If we work to leading and next-to-leading order in the number density,
calculating the energy exchange between plasma components will then
involve keeping a tally only of the kinetic energy, as in
(\ref{Ebkin}). This is because the potential energy is higher order in
the number density [or more precisely, the potential energy is higher
order in the plasma coupling $g$, to be defined later in
(\ref{gdefA})]. As the system interacts with other plasma components
through mutual Coulomb interactions, it will loose or gain energy
depending on the temperature gradients with other species, and the
energy exchange rate is given by
\begin{eqnarray}
  \frac{d {\cal E}}{dt} 
  = 
  \int d^3 v \,\frac{1}{2}\, m v^2 ~
  \frac{\partial f}{\partial t}({\bf v},t) \ .
\label{dedtbe}
\end{eqnarray}
\vskip0.1cm 
\noindent
In contrast to (\ref{dedtei}) and (\ref{dedteI}), for ease of notation
I have temporarily dropped the plasma component subscripts on the
rate, and I will keep with this convention until the final calculation
presented in Section~\ref{sec:calcleading}. We see that the entire
problem is bound up in calculating the rate of change $\partial
f/\partial t$ from an appropriate kinetic equation that captures the
relevant physics.  As it turns out, however, there is a serious
problem in performing all such calculation with the Coulomb potential
in three dimensions: the integrals in the kinetic equations diverge
logarithmically, and they do so at both large and small
distances.\footnote{\footnoteskip
  It is curious that this problem occurs for the Coulomb potential
  only in three dimensions, the case of most physical relevance in
  plasma physics.
}

For processes in which large-angle scattering is important, such as
the charged particle stopping power, it is natural to use the
Boltzmann equation, which I will write in the abbreviated form
\begin{eqnarray}
  \frac{\partial f}{\partial t} + 
  {\bf v} \! \cdot \!{\bm\nabla} f = B[f] \ ,
\label{BEsimp}
\end{eqnarray}
where ${\bm\nabla}$ is the gradient in position space, and $B[f]$ is
the scattering kernel, whose precise form will not concern us until
the next lecture.  The gradient vanishes because of spatial
uniformity, so we will set ${\bf v} \cdot {\bm\nabla} f=0$. The
Boltzmann equation was designed to account for the statistical effects
of short-distance collisions, and although its original context was
classical, quantum two-body scattering effects can easily be
incorporated. In fact, since the scattering phase shifts $\delta_\ell$
are known analytically for the Coulomb potential, Ref.~\cite{bps} used
this to calculate the two-body quantum corrections to all orders [in
the quantum parameter $\eta$ to be defined in (\ref{etadef})].  The
kernel $B[f]$ therefore contains all short-distance or ultraviolet
physics, for both classical and quantum scattering. However, (in three
spatial dimensions) the Coulomb potential is long-range, and the
integrals in $B[f]$ diverge in the infrared; or equivalently, if we
write the scattering kernel $B[f]$ in terms of momentum integrals, the
divergence appears at small values of momentum.  This was not a
problem in Boltzmann's original formulation of (\ref{BEsimp}), since
the Coulomb potential was unknown at that time, and he modeled
particle collisions in terms of hard-sphere scattering. In summary,
the Boltzmann equation gets the short-distance physics correct,
including quantum two-body scattering, but it misses the infrared
physics. The fact that the Boltzmann equation misses the long-distance
physics manifests itself as an infrared divergence in the scattering
kernel $B[f]$, thereby rendering calculations meaningless (unless we
tame, or regularize, this divergence).

Given that the Boltzmann equation misses the long-distance or infrared
(IR) physics, we might be tempted to try the Lenard-Balescu equation,
which I will write in the abbreviated form
\begin{eqnarray}
  \frac{\partial f}{\partial t} +
  {\bf v} \! \cdot \!{\bm\nabla} f  = L[f] \ ,
\label{LBEsimp}
\end{eqnarray}
where $L[f]$ is a scattering kernel whose exact form will be needed
only in the next lecture. Again, the gradient term will be set to
zero because of spatial uniformity. The Lenard-Balescu equation takes
the form of a Fokker-Plank equation, with the kernel $L[f]$ chosen to
capture the correct IR physics. However, for the Coulomb potential (in
three spatial dimensions), the Lenard-Balescu equation misses the
short-distance or ultraviolet (UV) physics, and this is manifested by
a UV divergence in $L[f]$. The situation for the Lenard-Balescu
equation is exactly reversed compared to that of the Boltzmann
equation. This is what Ref.~\cite{aono} calls the ``complementarity''
of these two kinetic equations, and in the dimensional continuation 
procedure we will use this fact to our advantage.

\section{Traditional Methods}

\subsection{Heuristic Models}
\label{sec:heuristic}

The rate equation (\ref{dedtbe}) reduces to a one-dimensional integral
over the entire range of physical length scales (or momentum scales,
if one so chooses), from zero all the way to infinity. Trouble arises
for the Coulomb potential in three dimensions since the integral in
question is logarithmically divergent at both integration limits. We
must therefore regulate the integral in some manner. Dimensional
continuation is one such procedure, but there are others. This
divergence problem was first worked around by simply cutting off the
divergent integrals by hand, with the cutoffs themselves being chosen
by physical arguments~(rather than a calculation).

The energy exchange rate we are considering is but an example of a
larger class of problems involving characteristic, but disparate,
length or energy scales in which the measured quantity of interest is
(logarithmically) insensitive to the physics above and below these
scales.  For these problems, the simplest and most intuitive
regularization scheme is to replace the offending integration limits,
{\em i.e.} infinity and zero, by the finite and non-zero physical
scales of the problem.  These two scales then act as formal
integration cutoffs, giving a finite logarithm of the ratio of the
scales. Furthermore, because the system is insensitive to the physics
above and below the respective cutoffs, this procedure provides a
physically meaningful result. Expressed in terms of length, we will
denote the long- and short-distance scales by $b_\text{max}$ and
$b_\text{min}$, and the integral over scales leads to a finite
logarithm involving the ratio of the physical length scales, so that
\begin{eqnarray}
  \frac{d{\cal E}}{dt} = K \, 
  \ln\!\left\{\frac{b_\text{max}}{b_\text{min}}\right\} \ ,
\label{ratemodel}
\end{eqnarray}
where $K$ is an easily determined prefactor with dimensions of energy
density per unit time. As we shall see, a calculation to leading order
in the number density is sufficient to provide the coefficient $K$,
while a next-to-leading order calculation is required to find the exact
terms under the logarithm.

The problem with this regularization prescription, which I will call
the {\em heuristic scheme}, is that we can only estimate the values of
the physical scales $b_\text{max}$ and $b_\text{min}$ to within
factors of order one or so. For example, it is physically reasonable
that the scale of the long distance cutoff in a plasma is set by a
Debye length, so that $b_\text{max}=c\,\kappa_\smD^{-1}$ with $c$
being a dimensionless constant of order unity; but what determines the
exact value of $c$\,?  In fact, how does \hbox{$b_\text{max} \sim
\kappa_\smD^{-1}$} arise naturally from the kinetic equations
themselves, rather than simply being chosen by hand? And should one
use the total Debye wave number $\kappa_\smD$, or just the
contribution from the electrons $\kappa_e$?  The origin of the short
distance cutoff $b_\text{min}$ is even less clear. In the extreme
classical limit, we expect this scale to be set by the classical
distance of closest approach $r_\text{cl}$ between two colliding
particles, so that \hbox{$b_\text{min} \sim
r_\text{cl}$}.\footnote{\footnoteskip 
  A further ambiguity arises in precisely defining $r_\text{cl}$, as
  this involves a somewhat arbitrary choice in the thermal averaging
  procedure.
} 
On the other hand, we expect quantum effects to dominate when
$r_\text{cl}$ becomes smaller than some thermal Compton wavelength
$r_\text{qm} \sim \hbar/q$, with $q$ being a typical thermally
averaged momentum transfer. In this case we expect $b_\text{min} \sim
r_\text{qm}$.  Worse yet, the intermediate region in which neither
classical nor quantum processes dominate is often realized in a weakly
coupled plasma, and in this case it is even more obscure how one
should choose $b_\text{min}$. We must interpolate between the extreme
classical and the extreme quantum scales, but in the literature the
exact procedure is always somewhat {\em ad hoc}. For example, a common
choice is to define $b_\text{min}^2$ to be the sum of the squares of
the classical and quantum scales, so that
\begin{eqnarray}
  b_\text{min}
  =
  \Big(r_\text{cl}^2 \,+\, r_\text{qm}^2\Big)^{1/2} \ .
\label{bminint}
\end{eqnarray}
I will have more to say about this in Section~\ref{sec:ginside}.

The heuristic scheme forces us to choose the specific forms of
$b_\text{max}$ and $b_\text{min}$ motivated by imprecise physical
arguments or heuristic exercises, which leads us into the art of model
building rather than systematic calculation. Indeed, the very notation
that we must {\em choose} a cutoff is misleading, since the physics
itself must conspire to render all integrals finite. Consequently, the
heuristic method suffers from an unknown coefficient under the
logarithm, and only the approximate value of the ratio $b_\text{max}/
b_\text{min}$ can be determined with this method (in fact, this ratio
varies across an order of magnitude over models in the literature,
rather than factors of two or three).  As we shall see, determining
the constant under the logarithm exactly is equivalent to determining
the next-to-leading order term exactly; therefore, models of the form
(\ref{ratemodel}) are accurate only to leading order, and no better.

\subsection{Convergent Kinetic Equations: Traditional Approach}

Rather than merely regulating the {\em integrals} in a rate derived
from the kinetic equations, as with (\ref{ratemodel}), a more
sophisticated approach involves regularizing the divergences in the
{\em kinetic equations} themselves.  In other words, the {\em theory}
itself is regularized, rather than a particular quantity being
calculated within the theory. The method of dimensional continuation
falls into this category, albeit with somewhat more subtle
mathematical machinery than traditional approaches. These approaches,
of which Refs.~\cite{fri,wei,gould} are good examples, are summarized
and placed into a common framework by Aono in Ref.~\cite{aono}.  As
discussed in Section~\ref{sec:problemwithproblem} of this lecture, one
can view the Boltzmann and Leonard-Balescu equations as providing
complementary physics since they both succeed and fail in
complementary regimes. The Boltzmann equation gets the short-distance
physics correct, while the Leonard-Balescu equation captures the
long-distance physics; conversely, Boltzmann and Leonard-Balescu miss
the long- and short-distance physics, respectively. This
complementarity motivates a class of kinetic equations of the
form~\cite{aono}
\begin{eqnarray}
  \frac{\partial f}{\partial t} + {\bf v} \! \cdot \!{\bm\nabla} f  
  = 
  B[f] + L[f] - R[f] \ ,
\label{convKE}
\end{eqnarray}
where $R[f]$ is a carefully chosen ``regulating kernel'' designed to
subtract the long-distance divergence of the Boltzmann equation and
the short-distance divergence of the Lenard-Balescu equation. At the
same time, the kernel $R[f]$ must preserve the correct short-distance
physics of the Boltzmann equation and the correct long-distance
physics of the Lenard-Balescu equation (a minimal requirement of the
regulating kernel $R[f]$ is that it take a Hippocratic Oath to {\em do
no harm}, at least to subleading order in the plasma coupling). Each
term on the right-hand-side of (\ref{convKE}) is separately divergent,
but collectively they lead to a finite collision kernel if properly
interpreted.\footnote{\footnoteskip 
  By ``properly interpreted'' I really mean that each term on the
  right-hand-side of (\ref{convKE}) should be separately regularized
  in some manner, rendering the individual kernels $B[f]$, $L[f]$, and
  $R[f]$ separately finite: $B[f]$ must be regulated at
  long-distances, $L[f]$ at short-distances, and $R[f]$ at both long-
  and short-distances. If this procedure is executed correctly, so
  that the long-distance divergences of $R[f]$ and $B[f]$ cancel, as
  do the short-distance divergences of $R[f]$ and $L[f]$, both in a
  consistent fashion, then the entire right-hand-side of
  (\ref{convKE}) remains finite as the cutoffs are removed. If $R[f]$
  does not disturb the ${\cal O}(g^2)$ physics, then the convergent
  kinetic equation will be accurate to ${\cal O}(g^2)$. I will have
  more to say about how one regulates long- and short-distances 
  {\em consistently} in Section~\ref{sec:lamb}, and how this can be
  a quite non-trivial process. 
}  
As an example, Gould and
DeWitt~\cite{gould} regulated the long-distance divergence of the
Boltzmann equation by simply replacing the Coulomb potential with a
Debye screened Yukawa-like potential,
\begin{eqnarray}
  V_\text{debye}(r)= 
  \frac{e^2}{4\pi r}\,\mathfrak{e}^{-\kappa_\smD\, r} \ ,
\label{Vscreen}
\end{eqnarray}
where, as not to confuse symbols, I write $\mathfrak{e}$ as the base
of the natural logarithm and $e$ as the electric charge. While
Ref.~\cite{gould} performed this operation by hand using physical
arguments, one could easily introduce a kernel $R[f]$ to do the same.

While the approach to convergent kinetic equations described by Aono
might appear to be more rigorous than the aforementioned model
building approach of (\ref{ratemodel}), it is no more systematic:
methods based on (\ref{convKE}) or its equivalent do not contain the
ability to estimate their own error, {\em i.e.} they cannot determine
their domain of applicability.  There is nothing in the formalism of
(\ref{convKE}) that keeps track of the plasma coupling constant, or
the order to which we are working in this constant. Indeed, one does
not generally think of (\ref{convKE}) in terms of a perturbation
theory.  In contrast, Ref.~\cite{bps} calculates the rate using a
systematic expansion in the plasma number density, or more precisely,
in a dimensionless plasma coupling parameter $g$ [to be defined by
(\ref{gdefA}) and discussed at length in Section~\ref{sec:eerate}]. 
Although written in a disguised form, the BPS rate coefficient
(\ref{bpsrate}) is an expansion to leading and next-to-leading order
in the coupling parameter: the leading order term goes like $g^2 \ln
g$, the next-to-leading order is proportional to $g^2$, and the ${\cal
O}(g^3)$ term provides an estimate of the error. Translating the work
of Gould and DeWitt~\cite{gould} into the language of a perturbative
expansion in a plasma coupling constant, it turns out that their
result is valid to order $g^2$ and is in agreement with BPS to this
order.\footnote{\footnoteskip
  For a detailed treatment of Gould and DeWitt in the context of BPS,
  see Appendix~B of Ref.~\cite{bps}.
}
However, in their final result, Gould and DeWitt retain spurious
higher order terms in $g^3$. I call these terms ``spurious'' because
Ref.~\cite{gould} did not calculate the full set of order $g^3$ terms,
but only some of these terms. Indeed, the notion of a systematic
expansion in a small dimensionless parameter does not enter their
calculational framework.  As this example shows, convergent kinetic
equations can be more accurate than the heuristic model building
technique of the previous section, but one cannot be sure of their
accuracy until a comparison with a systematic calculation has been
made, as with BPS and Ref.~\cite{gould}. 

\section{Bessel Function Example}
\label{sec:besselex}

\subsection{Analogy with Dimensional Continuation}

I will illustrate the main points of dimensional continuation with an
example involving the modified Bessel function $K_0(x)$, with an
emphasis on analytic continuation and how this can be used to extract
leading and next-to-leading order behavior. This example was first
presented in Ref.~\cite{lfirst}, and for pedagogical purposes it was
also included in Appendix~A of Ref.~\cite{bps}. This example contains
all the essential features of dimensional continuation, but in a
mathematically simple form, and while it is an imperfect analogy, as
all analogies are, it is explicit in all its details. We will show
that the modified Bessel function $K_0(x)$ has the expansion
\begin{eqnarray}
  K_0(x) 
  = 
  \underbrace{~-\ln x~~}_\text{LO} 
  ~+~\underbrace{~\ln 2-\gamma~}_\text{NLO}~
  +~{\cal O}(x^2) 
  =
  -\ln\!\left(\frac{e^\gamma}{2}\,x \right)
  ~+~{\cal O}(x^2) \ ,
\label{kzerolonlo}
\end{eqnarray}
to leading order (LO) and next-to-leading order (NLO) in $x$, with
$\gamma=0.577216 \cdots$ denoting Euler's constant. This expansion is
quite accurate for small values $x$, with an error of order $x^2$
rather than $x$ for symmetry reasons.  The asymptotic expansion
(\ref{kzerolonlo}) is a well known result~\cite{kzero}, but it is
rather difficult to prove by conventional methods because of the
non-analytic leading-log behavior. However, the method of dimensional
continuation allows us to derive this result rather easily. The price
one pays for this ease of derivation is that one must learn (or
recall) a bit of mathematical machinery which, at first sight, seems
unrelated to the problem at hand.

We start with the general integral representation of the modified 
Bessel functions \cite{knuint},
\begin{eqnarray}
  K_\nu(x) = \frac{1}{2}\int_0^\infty \! dk\,
  k^{\nu-1}\, \exp\left\{ -\frac{x}{2}\left(
  k + \frac{1}{k}\right)\right\} \ .
\label{knuint}
\end{eqnarray}
As the notation in (\ref{knuint}) suggests, we can think of $\nu$ as
the dimension of space and the integration variable $k$ as the wave
number. In this analogy, the argument $x$ corresponds to the
dimensionless coupling parameter $g$ of the plasma.  The following
dictionary provides a useful mnemonic in relating this mathematical
example to the plasma physics problem of real interest:
\begin{eqnarray}
\nonumber
  x &\rightarrow& \text{plasma coupling } g
\\[-6pt]
\nonumber
  \nu &\rightarrow& \text{spatial dimension}
\\[-6pt]
\nonumber
  k   &\rightarrow& \text{wave number}
\\[-6pt]
  dk\,k^{\nu-1}   &\rightarrow&  \text{integration measure }
  d^\nu k \ .
\label{physanalogy}
\end{eqnarray}
Pushing our physical analogy further, if we think of $\nu$ as being
the dimension of space, then it should always be a positive integer,
which I will express by the conventional set theory notation $\nu \in
\mathbb{Z}^+$; however, nothing \hbox{ {\em per se}} in the integral
representation (\ref{knuint}) requires that $\nu \in \mathbb{Z}^+$.
We can therefore think of $\nu$ in expression (\ref{knuint}) as being
a continuous real variable~\hbox{($\nu \in \mathbb{R}$)}, or indeed, a
complex variable \hbox{($\nu \in \mathbb{C}$)} if circumstances
warrant.  Similarly, for a real physical system written in the
appropriate integral form, there is nothing in any law of physics that
prevents us from interpreting the dimension of space $\nu$ as being a
complex number. Continuing $\nu$ from the positive integers into the
complex plane is what I mean by ``dimensional continuation.''  There
will be times when we restrict our attention to the real numbers only,
rather than the complex numbers in general, and I will refer to this
as dimensional continuation as well. As we shall see, this procedure
of taking $\nu \in \mathbb{R}$ or $\nu \in \mathbb{C}$ will allow us
to regulate otherwise infinite integrals in a systematic and
perturbative fashion.  Finite manipulations can then be performed, the
divergent poles will cancel from physically measurable quantities, and
afterward we can take $\nu$ to the appropriate integer dimension (in
this analogy we take $\nu \to0$, rather than $\nu \to 3$ as we do for
the physics problem).

\subsection{Leading Order Terms}

Let us first calculate the leading order in $x$ behavior of $K_\nu(x)$
for positive and negative values of $\nu$.  For small positive values
of $x$, the leading order $x$-behavior can be obtained by replacing
the exponential in (\ref{knuint}) by one, {\em except} in the regions
$k \to 0$ and $k \to \infty$, where the exponential is required for
convergence. Taking $\nu < 0$ first, note that the integral
(\ref{knuint}) is dominated by small values of $k$ near the lower
limit of integration. In terms of the analogy (\ref{physanalogy})
where $k$ is a wave number, this corresponds to the situation in which
long-distance IR physics is dominant.  Therefore, when $\nu < 0$ and
the integral is dominated by small values of $k$, the leading order
contribution to (\ref{knuint}) can be obtained from the leading order
behavior of the exponential, that is to say, the replacement
\begin{eqnarray}
  \exp\left\{ -\frac{x}{2}\left(
  k + \frac{1}{k}\right)\right\} 
  &\to&  
  \exp\left\{ -\frac{x}{2\, k} \right\} 
  ~~
  \text{with}
  ~~ 
  \vert x \vert \ll 1 ,~~\nu\!<\!0 
\label{expa}
\end{eqnarray}
will capture the entire leading order in $x$ behavior for negative
values of $\nu$. Note that (\ref{expa}) provides convergence as $k \to
0$, while large-$k$ convergence is provided by the prefactor
$k^{\nu-1}$ since $\nu < 0$. We will denote this leading order
contribution by $K_\nu^\smLT(x)$, and using the substitution
(\ref{expa}) we write
\begin{eqnarray}
  K_\nu^\smLT(x)
  &=& 
  \frac{1}{2}\int_0^\infty \! dk\,k^{\nu-1}\, 
  e^{- x/2k} 
  =
  \frac{1}{2}\left(\frac{x}{2}\right)^\nu \Gamma(-\nu) \ .
\label{knult}
\end{eqnarray}
In the last equality of (\ref{knult}), we have made the variable
change $y=x/2 k$, and we have used the standard integral
representation for the Gamma function,
\begin{eqnarray}
  \Gamma(z)
  = 
  \int_0^\infty \! dy \,y^{z-1}\, e^{-y} \ .
\label{gammaone}
\end{eqnarray}
This representation of the Gamma function only converges when $z > 0$
(actually, (\ref{gammaone}) is valid over the complex $z$-plane with
${\rm Re}\, z >0$). When $z=-\nu$ and $\nu<0$, we may indeed use
(\ref{gammaone}).

We can find the leading order in $x$ contribution when $\nu>0$ in a similar
manner. In this case, the integral is dominated by large
values of $k$ at the upper limit of integration, and we make 
the substitution
\begin{eqnarray}
  \exp\left\{ -\frac{x}{2}\left(
  k + \frac{1}{k}\right)\right\} 
  &\to&  
  \exp\left\{ -\frac{x\, k}{2}\right\} 
  \hskip1cm 
  \text{with}~~ \vert x \vert \ll 1 \ , ~\nu > 0 \ ,
\label{expb}
\end{eqnarray}
thereby giving the leading order result
\begin{eqnarray}
  K_\nu^\smGT(x)
  &=& 
  \frac{1}{2}\int_0^\infty \! dk\,k^{\nu-1}\, e^{- x k/2}
  =
  \frac{1}{2}\left(\frac{x}{2}\right)^{-\nu} \Gamma(\nu) \ .
\label{knugt}
\end{eqnarray}
Note that the exponential provides convergence as $k \to \infty$,
while the integrand possesses an integrable singularity at $k=0$ 
for $0<\nu<1$ (the integrand is non-singular at $k=0$ when $\nu 
\ge 1$). 

We will eventually take the $\nu \to 0$ limits of (\ref{knult}) and
(\ref{knugt}), since we are interested in $K_0(x)$ and not
$K_\nu(x)$. While we could take $\nu$ to be a general real or complex
number until the limit is required, it is easier to consider only
small values of $\nu$ from the start (we need work no higher than
linear order, since this and higher orders vanish when $\nu\to 0$).
To do this, we expand the Gamma function $\Gamma(\pm\nu)$ to
linear-order in its argument using $\Gamma(z)=1/z - \gamma + {\cal
O}(z)$. Taking $z=\pm \nu$ in this expansion gives
\begin{eqnarray}
  K_\nu^\smGT(x)
  &=& 
  \phantom{-}
  \frac{1}{2\nu}\, \left(\frac{x}{2}\right)^{\!-\nu}
  \Big[1 - \nu \gamma \Big] 
  \hskip1cm : ~\text{LO in $x$ when } \nu > 0 
\label{knua}
\\[5pt]
  K_\nu^\smLT(x)
  &=& 
  -\frac{1}{2\nu}\, \left(\frac{x}{2}\right)^\nu~
  \Big[1 + \nu \gamma \Big] 
  \hskip1.05cm : ~\text{LO in $x$ when } \nu < 0 \ .
\label{knub}
\end{eqnarray}
Expressions (\ref{knua}) and (\ref{knub}) are accurate to linear order
in the dimension $\nu$; on the other hand, (\ref{knua}) gives the
leading order in $x$ contribution to $K_\nu(x)$ as defined by
(\ref{knuint}) when $\nu>0$, and (\ref{knub}) gives the leading order
in $x$ contribution to $K_\nu(x)$ when $\nu<0$. To compare these two
expressions, we must analytically continue them to a common
dimension. We will discuss this further in the next section.

In terms of our physics analogy, expression (\ref{knua}) captures the
leading order short-distance physics in the $\nu> 0$ regime; the pole
at $\nu = 0$ corresponds to a small-$k$ divergence, which, pushing our
physical analogy again, would reflect missing or incomplete
long-distance physics (as with the Boltzmann equation). The situation
is completely reversed for (\ref{knub}), which captures the leading
order long-distance physics for $\nu> 0$, with the pole at $\nu=0$
corresponding to a large-$k$ divergence arising from missing
short-distance physics (like the Lenard-Balescu equation). As
functions of $\nu$, we see from (\ref{knua}) and (\ref{knub}) that
$K_\nu^\smGT(x)$ and $K_\nu^\smLT(x)$ are analytic in $\nu$, except
for the simple pole at $\nu=0$. As we shall see, the analytic
continuation to complex $\nu$ takes the same functional form as the
individual expressions (\ref{knua}) and (\ref{knub}), each defined
separately for $\nu>0$ and $\nu<0$, respectively.

\subsection{Some Comments on Analytic Continuation}
\label{sec:analytic}

Since analytic continuation plays such a central role in dimensional
continuation, at least mathematically, I would like to briefly discuss
the conditions under which a function can be analytically continued
from one region of the complex plane to another. Recall that a
function $f$ is said to be {\em analytic at a point} $z_0$ in the
complex plane ${\mathbb C}$, if and only if its derivative exists not
only at $z_0$, but also at every point within some open neighborhood
of $z_0$. A function $f$ is {\em analytic on a domain} $D$ in the
complex plane if it is analytic at each point in~$D$. Analyticity is a
very stringent condition on a function, since the existence of the
derivative of a complex function is a much more robust constraint than
the corresponding existence of the derivative of a function on a real
domain. This is because in the two-dimensional complex plane, the
limiting procedure defining the derivative must exist regardless of
the direction used in taking the limit.  In fact, analyticity at a
point $z_0$ is so strong that it implies the existence and continuity
of all derivatives $f^{(n)}(z_0)$ for any order $n > 0$
\cite{cbvA}. In other words, an analytic function on $D$ can be
thought of as being infinitely smooth on $D$, even though the
definition of analyticity itself invokes only the existence of the
first derivative, albeit on a neighborhood.

Analyticity is such a stringent condition, that the behavior of an
analytic function in a small domain is enough to determine its
behavior in a larger region. Even if the function is only known along
a one-dimensional curve in the complex plane, such as a portion of the
real axis, this is enough to uniquely determine the function in the
complex plane.\footnote{\footnoteskip
  In fact, Carlson's Theorem~\cite{Carlsonth} can be used to uniquely
  extend a function defined only on the integers to the whole complex
  plane. This is actually the theorem of most relevance here; however,
  as we shall see in Section~\ref{sec:math}, in practice we can
  analytically continue a function defined on the integers without
  this Theorem. We appeal to Carlson's Theorem only to guarantee the
  uniqueness of this procedure.
} 
This is because the derivatives of an analytic function exist to
all orders, and therefore a Taylor series expansion about a point of
analyticity always exists with some non-zero radius of
convergence. For example, one of the powers of analytic continuation
is that the original function may be defined in any manner over a
subregion, and this can generate a unique function over a larger
region. As an application of this, take the analytic function defined
by the infinite geometric series
\begin{eqnarray}
  f(z) = \sum_{n=0}^\infty z^n \ ,
\label{fseriesdef}
\end{eqnarray}
a series that converges only for $\vert z \vert < 1$. Upon defining
$f$ by (\ref{fseriesdef}), we therefore take the domain $\mathbb{D}_1$
to be the unit disk about the origin, excluding the unit circle
itself.  In this domain, the geometric series converges to
\begin{eqnarray}
  \sum_{n=0}^\infty z^n = \frac{1}{1-z} 
  ~~~\text{for}~~ z \in \mathbb{D}_1 \ .
\end{eqnarray}
Note, however, that the function $g(z)=1/(1-z)$ is defined over the
entire complex plane except $z=1$, a region I will call
$\mathbb{C}_1$.  Since the function $f$ is only defined within the
unit circle, and since $f$ and $g$ agree within the unit circle,
the function $g\!:\!  \mathbb{C}_1 \to \mathbb{C}$ is the unique
analytic continuation of $f\!:\!D_1 \to \mathbb{C}$. 

As a more relevant example, consider $K_\nu^\smLT(x)$ with $\nu < 0$
as given by (\ref{knub}). To compare this with $K_\nu^\smGT(x)$, which
is determined by (\ref{knua}) for $\nu>0$, we must analytically
continue $K_\nu^\smLT(x)$ to the positive real $\nu$-axis.  We can
think of $K_\nu^\smLT(x)$ as a sequence of functions of an independent
variable $\nu$ indexed by a continuous label $x$; therefore, in a more
suggestive notation, we temporarily write $f_x(\nu) \equiv
K_\nu^\smLT(x)$\,. While the collection of functions $f_x(\nu)$ are
defined in (\ref{knub}) on the negative $\nu$-axis $\mathbb{R}^-$
(excluding the simple pole at zero), they can be analytically
continued to the complex $\nu$-plane $\mathbb{C}$. Furthermore, the
functions take the same algebraic form on the complex plane, namely,
\begin{eqnarray}
  f_x(\nu)=-\frac{1}{2\nu}\,\left(\frac{x}{2}\right)^\nu 
  \Big[1 + \nu\, \gamma \Big]
  ~~\text{for}~~ \nu \in \mathbb{C} \ .
\label{fxnu}
\end{eqnarray}
We can now restrict our attention from $\mathbb{C}$ in general to the
positive $\nu$-axis $\mathbb{R}^+$ (excluding zero). This allows us to
directly compare $K_\nu^\smLT(x)$ and $K_\nu^\smGT(x)$ at $\nu \in
\mathbb{R}^+$ using the same algebraic forms as given by (\ref{knub})
and (\ref{knua}).  In the next section, we discuss the implications of
analytically continuing $K_\nu^\smLT(x)$ from the negative axis
$\nu<0$ to the positive axis $\nu>0$. Alternately, we could continue
$g_x(\nu) \equiv K_\nu^\smGT(x)$ to the region $\nu \in \mathbb{R}^-$
using the same functional form as (\ref{knua}), and compare this with
$K_\nu^\smLT(x)$.

\subsection{Next-to-Leading Order Term}
\label{sec:knunlo}

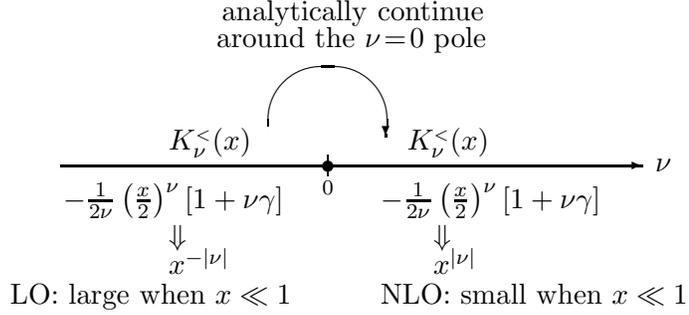
\begin{figure}[t]
\begin{picture}(120,120)(-75,0)
\put(-101,60){\vector(1,0){220}}
\put(124,58){$\nu$}

\put(0,56){\line(0,1){8}}
\put(-2.65,57.3){$\bullet$}
\put(-2.00,49){${\scriptstyle 0}$}

\put(-60,66){$K_\nu^\smLT(x)$}
\put(-100,44){$-\frac{1}{2\nu}\left(\frac{x}{2}
  \right)^\nu[1 + \nu\gamma]$}
\put(-60,30){$\Downarrow$}
\put(-60,19){$x^{-\vert \nu \vert}$}
\put(-120,8){LO: large when $x \ll 1$}

\put(30,66){$K_\nu^\smLT(x)$}
\put(20,44){$-\frac{1}{2\nu}\left(\frac{x}{2}
  \right)^\nu[1 + \nu\gamma]$}
\put(40,30){$\Downarrow$}
\put(40,19){$x^{\vert \nu \vert}$}
\put(20,8){NLO: small when $x \ll 1$}

\put(0,75){\oval(45,45)[t]}
\put(22,72){\vector(0,-1){1}}
\put(-40,115){analytically continue}
\put(-42,105){around the $\nu\!=\!0$ pole}

\end{picture}
\caption{\captionskip The analytic continuation of $K_\nu^\smLT(x)$
from $\nu\in{\mathbb R}^-$ to $\nu\in{\mathbb R}^+$ in the complex
$\nu$-plane: the same expression can be used for $K_\nu^\smLT(x)$
throughout the complex plane since the pole at $\nu=0$ can easily be
avoided, as indicated in the figure. Note that $K_\nu^\smLT(x) \sim
x^{-\vert \nu \vert}$ is leading order in $x$ for $\nu < 0$. However,
upon analytically continuing $K_\nu^\smLT(x)$ to $\nu>0$, the
$x$-dependence becomes $K_\nu^\smLT(x) \sim x^{\vert \nu \vert}$, and
the function is next-to-leading order relative to $K_\nu^\smGT(x) \sim
x^{-\vert \nu \vert}$\,.  }
\label{fig:knucont}
\end{figure}

We now illustrate the key mathematical result that allows dimensional
continuation to extract not only the leading, but the next-to-leading
order terms. Recall from (\ref{knua}) and (\ref{knub}) that
$K_\nu^\smGT(x)$ and $K_\nu^\smLT(x)$ are both leading order in $x$
for $\nu>0$ and $\nu<0$, respectively. Since these functions were
calculated for mutually exclusive values of $\nu$, one might think
that they cannot be compared. When viewed as an analytic function in
the complex \hbox{$\nu$-plane}, however, we have seen that
$K_\nu^\smLT(x)$ is also a function over the domain $\nu>0$, in which
case both $K_\nu^\smLT(x)$ and $K_\nu^\smGT(x)$ can be compared at the
same values of $\nu$ and $x$.  Since the algebraic form is so simple,
$K_\nu^\smLT(x)$ takes the same functional form when analytically
continued to $\nu>0$ as it did for $\nu<0$. As illustrated in
Fig.~\ref{fig:knucont}, this means that $K_\nu^\smLT(x)$ becomes
next-to-leading order in $x$ along the positive real axis:
\begin{eqnarray}
  K_\nu^\smGT(x)
  &=& 
  \phantom{-}
  \frac{1}{2\nu}\, \left(\frac{x}{2}\right)^{\!-\nu}
  \Big[1 - \nu \gamma \Big] 
  \hskip1cm : ~\text{LO in $x$ when } \nu > 0 
\label{knuaplus}
\\[10pt]
  K_\nu^\smLT(x)
  &=& 
  -\frac{1}{2\nu}\, \left(\frac{x}{2}\right)^\nu~
  \Big[1 + \nu \gamma \Big] 
  \hskip1.04cm : ~\text{NLO in $x$ when } \nu > 0 \ .
\label{knubplus}
\end{eqnarray}
The expressions (\ref{knuaplus}) and (\ref{knubplus}) remain accurate
to linear-order in $\nu$. 

To see that (\ref{knubplus}) is indeed next-to-leading order in $x$
relative to (\ref{knuaplus}), note that the \hbox{$x$-behavior} of the
leading order contribution for $\nu>0$ can be written
\hbox{$K_\nu^\smGT(x) \sim x^{-\vert \nu\vert}$}. I have used the
absolute value $\vert \nu \vert$ to emphasize that the power of $x$ in
(\ref{knuaplus}) is strictly negative when $\nu>0$.  Similarly, along
the positive $\nu$-axis we find the behavior $K_\nu^\smLT(x) \sim
x^{\vert \nu \vert}$ for $\nu>0$, and we see that $x^{-\vert \nu
\vert} \gg x^{\vert \nu\vert}$ for $0<x \ll1$. This means
$K_\nu^\smGT(x) \gg K_\nu^\smLT(x)$ for $\nu>0$ and $0<x\ll 1$, and we
are therefore justified in calling $K_\nu^\smGT(x)$ leading order in
$x$ and $K_\nu^\smLT(x)$ next-to-leading order.

Strictly speaking, we have only shown that $K_\nu^\smLT(x)$ is
subleading relative to $K_\nu^\smGT(x)$ when $\nu>0$. To conclude that
$K_\nu^\smLT(x) \sim x^\nu$ is indeed next-to-leading order relative
to $K_\nu^\smGT(x) \sim x^{-\nu}$, it is important to establish that
there are no powers of $x$ between $x^\nu$ and $x^{-\nu}$ in the
expansion of $K_\nu(x)$.  For $\nu>0$, one simply subtracts
(\ref{knua}) from (\ref{knuint}), and it becomes clear that this error
is higher order in $x$ than $x^\nu$.  For a more detailed proof of
this, see footnote~2 of Ref.~\cite{lfirst}. A similar statement holds
for $K_\nu^\smGT(x)$, namely, as we analytically continue from $\nu>0$
to $\nu<0$, the quantity $K_\nu^\smGT(x)$ switches from leading order
to next-to-leading order in~$x$ relative to $K_\nu^\smLT(x)$.

\subsection{Assembling the Pieces}
\label{sec:assembly}

We have now assembled enough results to find $K_0(x)$ to leading and
next-to-leading order in $x$: we simply add the expressions
(\ref{knuaplus}) and (\ref{knubplus}) and take the limit of vanishing
$\nu$. Note that this does not lead to any form of ``double
counting.'' Instead, we are simply adding the next-to-leading order
term (\ref{knubplus}) to the leading order term (\ref{knuaplus}) at a
common value of~\hbox{$ \nu > 0$}. Upon taking the limit of vanishing
$\nu$, or more precisely $\nu \to 0^+$ since $\nu$ is always positive
in (\ref{knuaplus}) and (\ref{knubplus}), we obtain $K_0(x)$ to
leading and next-to-leading order in $x$.

We now calculate this limit, proving that 
\begin{eqnarray}
  \lim_{\nu \to 0^+}\Big[K_\nu^\smGT(x) + K_\nu^\smLT(x) \Big]
  =
  - \ln x + \ln 2 - \gamma \ .
\label{knuaa}
\end{eqnarray}
Let us first expand $(x/2)^{\pm \nu}$ in powers of $\nu$. We will
denote $z=x/2$, from which we find $z^{\pm \nu} = e^{\ln( z^{\pm
\nu})} = e^{\pm\nu \ln z} = 1 \pm \nu \ln z + {\cal O}(\nu^2)$, or in
summary:
\begin{eqnarray}
  z^{\pm \nu} 
  =
  1 \pm \nu \ln z + {\cal O}(\nu^2) \ .
\label{zpmnu}
\end{eqnarray}
This allows us to express (\ref{knuaplus}) and (\ref{knubplus}) as
\begin{eqnarray}
\label{knuGT}
  K_\nu^\smGT(x)
  &=& 
  \frac{1}{2\nu}\, 
  \bigg[ 1 - \nu\ln\!\left(\frac{x}{2}\right) \bigg] 
  \bigg[1 - \nu \gamma \bigg] + {\cal O}(\nu) 
  = 
  \phantom{-}
  \frac{1}{2\nu} - \frac{1}{2}\ln\!\left(\frac{x}{2}\right) -
  \frac{\gamma}{2}   + {\cal O}(\nu) \ ,
\end{eqnarray}
and 
\begin{eqnarray}
\label{knuLT}
  K_\nu^\smLT(x)
  &\!=\!& 
  -\frac{1}{2\nu}\, 
  \bigg[ 1 + \nu\ln\!\left(\frac{x}{2}\right) \bigg] 
  \bigg[1 + \nu \gamma \bigg] + {\cal O}(\nu) 
  = 
  -\frac{1}{2\nu} - \frac{1}{2}\ln\!\left(\frac{x}{2}\right) -
  \frac{\gamma}{2}   + {\cal O}(\nu)  .
\end{eqnarray}
When we divide (\ref{zpmnu}) by a factor of $\nu$, as required by
(\ref{knuaplus}) and (\ref{knubplus}), note that we find: (i) a pole
$1/\nu$ from the first term in (\ref{zpmnu}), (ii) a non-analytic
finite contribution $\pm\ln z$ from the second term, and (iii) the
error in $\nu$ becomes ${\cal O}(\nu)$, which is the same order in
$\nu$ that we are neglecting in (\ref{knuaplus}) and
(\ref{knubplus}). The error in $\nu$, which vanishes in the limit $\nu
\to 0^+$, should not be confused with the error in $x$, the latter
being ${\cal O}(x^2)$ for vanishing $\nu$. Note that the pole terms
cancel upon adding (\ref{knuGT}) and (\ref{knuLT}), so that
\begin{eqnarray}
  K_\nu^\smGT(x) + K_\nu^\smLT(x)
  &=& 
  - \ln\!\left(\frac{x}{2}\right) - \gamma + {\cal O}(\nu) \ ,
\label{loandnlo}
\end{eqnarray}
thereby giving (\ref{knuaa}) as $\nu \to 0^+$. As we have discussed,
there are no $x$-dependent terms that lie between $K_\nu^\smGT(x) 
\sim x^{-\nu}$ and $K_\nu^\smLT(x) \sim x^\nu$, so this procedure has
captured the leading and next-to-leading order behavior in $x$.

In exactly the same way, we can also calculate the leading order and
next-to-leading order contribution to $K_0(x)$ by taking the limit
from the left,
\begin{eqnarray}
  \lim_{\nu \to 0^-}\bigg[
  K_\nu^\smLT(x) + K_\nu^\smGT(x)  \bigg] 
  =
  - \ln x + \ln 2 - \gamma \ .
\label{knubb}
\end{eqnarray}
I should point out a potential notational problem in (\ref{knuaa}) and
(\ref{knubb}). Concentrating on (\ref{knuaa}) for the moment, the
limit $\nu \to 0^+$ indicates that both terms $K_\nu^\smGT$ and
$K_\nu^\smLT$ are understood to live in dimensions $\nu>0$, with the
second term $K_\nu^\smLT$ having been analytically continued from
\hbox{$\nu<0$}. The notation with which the term $K_\nu^\smLT$ is
written in (\ref{knuaa}), however, does not indicate that it has been
analytically continued. This should be no cause for confusion,
however, since $K_\nu^\smLT$ takes the same functional form in any
dimension $\nu$; therefore, a separate notation indicating that the
$K_\nu^\smLT$ in (\ref{knuaa}) has been analytically continued is
unnecessary. We can simply add $K_\nu^\smGT$ and $K_\nu^\smLT$ as
calculated in $\nu>0$ and $\nu<0$ respectively.

\pagebreak
\section{Dimensional Continuation}

\subsection{Rate of Energy Exchange as a Perturbative Expansion}
\label{sec:eerate}

\subsubsection{The $g$-Expansion}

Before moving on to the details of dimensional continuation, we must
first discuss the plasma expansion parameter $g$. Since the problems
in plasma physics are usually so complicated as to preclude a
perturbative approach, most plasma physicists do not usually think in
terms of expanding systematically in a small dimensionless parameter.
However, for a weakly to moderately coupled plasma, it is a quite
fruitful approach to perturbatively expand the rate in a small
dimensionless coupling constant.

That such a universal parameter for a plasma exists was discussed at
length in Ref.~\cite{by}, where it was shown that any physical
quantity associated with a plasma whose species are in equilibrium
with themselves (such as the plasma we are studying) can be expanded
in {\em integer} powers of a dimensionless coupling constant $g$
defined by
\begin{eqnarray}
\nonumber\\[-10pt]
\label{gdef}
  g 
  &\equiv&
  \frac{\text{Coulomb Energy for Two Charges at Separation } 
  \lambda_\smD}
  {\text{Temperature in Energy Units}} \ ,
\\[-10pt]\nonumber
\end{eqnarray}
with $\lambda_\smD$ being the Debye length of the plasma. Since the
potential energy between two like charges is given by $V=e^2/4\pi r$
in rationalized units, and writing the Debye wave number as
$\kappa_\smD=\lambda_\smD^{-1}$, the coupling parameter is
therefore
\begin{eqnarray}
  g 
  &=&
  \frac{e^2\,\kappa_\smD}{4\pi\, T} \ .
\label{gdefA}
\end{eqnarray}
For a multicomponent plasma, there is actually a coupling constant 
for each pair of components, 
\begin{eqnarray}
\label{gabdef}
 g_{ab} &=& \frac{e_a e_b\, \kappa_b}{4\pi T_b} \ ,
\end{eqnarray}
with $\kappa_b$ defined by (\ref{kbdef}). However, when the pairs have 
approximately the same coupling strength, then the single parameter 
(\ref{gdefA}) adequately characterizes the entire plasma. Expressing 
the charges as $e_a = Z_a\, e$ and $e_b = Z_b\, e$, we can write $g_{ab}$ 
as 
\begin{eqnarray}
  g_{ab}
  &=& 
  Z_a Z_b^2 \, \frac{e^3}{4\pi}\, \frac{n_b^{1/2}}{T_b^{3/2}} \ ,
\end{eqnarray}
and we see that the coupling constant is proportional to the cube of
the electric charge, the square root of the density, and the inverse
(3/2)-power of the temperature.

Recall that the usual plasma parameter $\Gamma$ is defined in a
similar manner to (\ref{gdef}), except that the charge separation is
determined not by $\lambda_\smD$, but by the inter-particle spacing in
the plasma,
\begin{eqnarray}
\nonumber\\[-10pt]
\label{gammadef}
  \Gamma 
  &\equiv&
  \frac{\text{Coulomb Energy for Two Charges at Inter-particle 
  Separation } d}{\text{Temperature in Energy Units}} \ .
\\[-10pt]\nonumber
\end{eqnarray}
The inter-particle spacing $d$ is defined in several ways in the
literature, but the idea is to transform the plasma number density $n$
into a length scale, so that $d \propto n^{-1/3}$. The most common
convention is to define $d$ to be the radius of a sphere containing,
on average, a single plasma particle, so that $4\pi\, d^3/3=1/n$, and
therefore
\begin{eqnarray}
  \Gamma 
  =
  \frac{e^2}{4\pi T}\, 
  \left(\!\frac{4\pi n}{3}\!\right)^{1/3} \ .
\end{eqnarray}
With this convention, the relation between the two plasma coupling
parameters for a single plasma species is $\Gamma^3 =g^2/3$, and for
an arbitrary number of plasma species we always find $g \propto
\Gamma^{3/2}$. We can therefore use either $g$ or $\Gamma$ to
characterize the strength of the plasma, as $g$ and $\Gamma$ become
large or small together.\footnote{\footnoteskip \label{foot:smallg}
  The dimensional continuation method requires that we perform
  calculations in an arbitrary spatial dimension $\nu$; however, the
  parameter $g=e^2 \kappa_\smD/4\pi T$ as given by (\ref{gdefA}) is
  {\em dimensionless} (in the engineering sense) only in three spatial
  dimensions. As such, it is meaningless to call $g$ large or small in
  any dimension other than $\nu\!=\!3$. This is no cause for alarm,
  however, since in footnote~\ref{foot:gnudef}, I will construct a
  dimensionless coupling parameter $g_\nu$ in arbitrary spatial
  dimensions $\nu$.  The parameter $g_\nu$ can then be used to
  characterize the plasma strength in $\nu$ dimensions, and it will
  have the properties that (i)~$g_\nu \propto g$ and (ii) $g_\nu \to
  g$ as $\nu \to 3$. Property (i) implies that a $g_\nu$-expansion is
  the same as a $g$-expansion, and (ii) implies this correspondence
  continues down to three dimensions, where $g$ becomes the relevant
  expansion parameter.
} For our purposes, however, there is an an advantage to using $g$
rather than $\Gamma$ since physical quantities are expanded as {\em
integer} powers of $g$, while they expand in fractional powers of
$\Gamma$.

\subsubsection{Next-to-Leading Order and the Coulomb Logarithm}

As I have said, any plasma quantity can be written as a power series
expansion in integer power of $g$, with the possible exception of
non-analytic terms involving $\ln g$. For the process of energy
exchange via Coulomb interactions, this non-analyticity arises from
the competition between disparate physical length scales.  As an
expansion in $g$, the rate of energy exchange takes the form
\begin{eqnarray}
  \frac{d{\cal E}}{dt} 
  = 
  -\underbrace{A\, g^2\ln g}_\text{LO}
  \,+\, 
  \underbrace{~B g^2~}_\text{NLO} \,+\,  {\cal O}(g^3) \ .
\label{dedtNLO}
\end{eqnarray}
In (\ref{dedtNLO}), I have indicated the leading order in $g$ (LO) and
the next-to-leading order in $g$ (NLO) terms in the
\hbox{$g$-expansion}: the first term is leading order relative to the
second because $\vert g^2\ln g\vert > g^2$ for small $g$. The minus
sign on the leading order term of (\ref{dedtNLO}) is a matter of
convention. Since the logarithm $\ln g$ will be negative in a weakly
coupled plasma (recall $g <1$), the minus sign renders the coefficient
$A$ positive when the energy exchange is positive. The coefficient $A$
was first calculated by Spitzer. The coefficient $B$, however, is very
difficult to calculate, and this was the main purpose of
BPS~\cite{bps}. It is convenient to define the dimensionless
coefficient $C$ by $B=-A\, \ln C$, in which case we can write
\begin{eqnarray}
  \frac{d{\cal E}}{dt} 
  &=& 
  A\, g^2\ln\Lambda_\smCoul \,+\, {\cal O}(g^3) \ ,
  ~\text{with}~~
  \ln\Lambda_\smCoul = -\ln\!\left\{ C g\right\} \ .
\label{lngsqu}
\end{eqnarray}
We see, then, that knowing the next-to-leading order term is
equivalent to knowing the exact coefficient $C$ under the
logarithm. Note that the minus sign renders the Coulomb logarithm
positive when $g$ is very small, in keeping with convention.

\subsubsection{Factors of $g$ Inside the Coulomb Logarithm}
\label{sec:ginside}

For the heuristic model building of Section~\ref{sec:heuristic}, let
us pause for a moment and show that the argument of the Coulomb
logarithm in (\ref{ratemodel}) is indeed proportional to $g$, as
required by (\ref{lngsqu}). On physical grounds we saw that the
long-distance scale $b_\text{max}$ is set by a Debye length, and
therefore we nominally set \hbox{$b_\text{max}= \kappa_\smD^{
-1}$}. In the extreme classical limit, the short-distance cutoff
$b_\text{min}$ is set by the classical distance of closest approach
$r_\text{cl}$, so that \hbox{$b_\text{min}=c \, r_\text{cl}$}. For
simplicity, we will choose the coefficient $c$ such that $K = A\,g^2$
between (\ref{ratemodel}) and (\ref{lngsqu}), in which case
\begin{eqnarray}
  \ln\Lambda_\smCoul 
  = 
  \ln\!\left\{ \frac{b_\text{max}}{b_\text{min}}\right\} \ .
\label{lnmodel}
\end{eqnarray}
Let us consider two unit charges of mass $m$ approaching one another
with zero impact parameter. The rms speed of each particle is
determined by
\begin{eqnarray}
  \frac{1}{2}\, m\, \bar v^2 = \frac{3}{2}\, T 
  ~~~ \Rightarrow ~~ \bar v = \sqrt{\frac{3\,T}{m}} \ ,
\label{vbardef}
\end{eqnarray}
while energy conservation $\frac{1}{2}\,m \bar v^2 + \frac{1}{2}\,m
\bar v^2 = e^2/4\pi r_\text{cl}$ gives the distance of closest approach,
\begin{eqnarray}
  r_\text{cl}
  = 
  \frac{e^2}{4\pi\, m \bar v^2}
  =
  \frac{e^2}{4\pi}\,\frac{1}{3 T} \ .
\label{rzero}
\end{eqnarray}
In the extreme classical regime, we see that
the argument of the Coulomb logarithm in (\ref{lnmodel})
is indeed proportional to the plasma coupling constant, 
\begin{eqnarray}
  \frac{b_\text{min}}{b_\text{max}}
  =
  c \,r_\text{cl}\, \kappa_\smD 
  =
  \frac{c}{3}\,\frac{e^2\, \kappa_\smD}{4\pi\, T} 
  =
  \frac{c}{3}\, g \ .
\end{eqnarray}

Let us now look at the {\em ad hoc} interpolation (\ref{bminint})
between the classical and quantum regimes.  Up to this point I have
said very little about quantum mechanics. While I will not dwell on
quantum corrections, I will briefly discuss a dimensionless expansion
parameter that characterizes the strength of the quantum two-body
scattering correction. There are many ways of defining such a
parameter, but I will follow Ref.~\cite{by}, taking
\begin{eqnarray}
\nonumber\\[-10pt]
  \eta 
  &\equiv&
  \frac{\text{Classical Distance of Closest Approach}}
  {\text{Thermal Wavelength}} \ .
\label{etadefwords}
\\[-10pt]\nonumber
\end{eqnarray}
With this definition, quantum corrections are large when $\eta \ll 1$.
Motivated by the de~Broglie wavelength of a particle, the thermal
wavelength of a plasma species is given by $r_\text{qm} = \hbar/\bar
q$, where $\bar q = m \bar v$ is a typical momentum transfer suffered
during a collision. Definition (\ref{etadefwords}) yields $\eta =
r_\text{cl}/r_\text{qm} = (e^2/4\pi m \bar v^2)\cdot (m \bar
v/\hbar)$, or more succinctly
\begin{eqnarray}
  \eta 
  =
  \frac{e^2}{4\pi \hbar\, \bar v}\ ,
\label{etadef}
\end{eqnarray}
from which (\ref{bminint}) gives
\begin{eqnarray}
  \frac{b_\text{min}}{b_\text{max}}
  =
  \frac{c}{3}\, \left(1 + \frac{1}{\eta^2} \right)^{1/2} \! g \ .
\end{eqnarray}
In the extreme quantum limit in which $\eta \ll 1$, this becomes
$b_\text{min}/b_\text{max}=(c/3)\,(g/\eta)$. 

Finally, note that the factors inside the BPS Coulomb logarithm
(\ref{bpsrate}) are also proportional to the coupling constant $g$,
and upon dropping the electron subscripts for convenience, we find
\begin{eqnarray}
  \frac{\hbar^2 \omega^2}{T^2} 
  = 
  \frac{1}{3}\,  \frac{g^2}{\eta^2} \ .
\label{hwot}
\end{eqnarray} 
We see that the $g$-dependence of the Coulomb logarithm arises
quite naturally. However, the accompanying coefficient under the
logarithm might also possess $\eta$-dependence, thereby obscuring
the $g$-dependence unless we are careful. 

\subsection{Mathematics of Dimensional Continuation}
\label{sec:math}

Before describing what dimensional continuation is, allow me to first
state what it is not. Dimensional continuation is {\em not} performing
an integral to a fractional power of the spatial dimension, as with
the meaningless expression
\begin{eqnarray}
\nonumber
  \int_0^\infty \!\! d^{\frac{3}{2}}k\, f(k) \ .
\end{eqnarray}
Instead, dimensional continuation is the following. Suppose some
physical quantity of interest can be written as an integral over a
kernel
\begin{eqnarray}
  Q(m)
  =
  \int_{{\mathbb R}^3} \! d^3 k\, f({\bf k}; m) \ ,
\label{intthree}
\end{eqnarray}
where $f$ is determined by the physical equations of motion, whether
classical or quantum. The integrand is of course a function of the
physical parameters, such as the masses and charges of the fundamental
particles, and I have abbreviated this dependence by the parameter
$m$. For definiteness, we will think of $k$ as a wave number with
dimensions of an inverse length. The laws of physics, from which
(\ref{intthree}) follows, are usually written in three dimensional
space. Thus, we usually take ${\bf k}$ to be a three-dimensional
vector, and we integrate ${\bf k}$ over the entire three-dimensional
Euclidean space ${\mathbb R}^3$.

The known fundamental laws of physics themselves, however, do not
specify a particular spatial dimension in which they hold. In fact, as
far as the known laws of physics are concerned, the actual value $\nu$
of the spatial dimension can be viewed as a free {\em integer}
parameter: it is simply an unexplained empirical fact that we live in
three dimensions.\footnote{\footnoteskip
  One would expect that the fundamental theory of nature, a theory of
  everything, would predict the number of space-time dimensions,
  solving this mystery at last. One of the great successes of string
  theory is that it is one of only two known theories that indeed
  predicts the number of space-time dimensions -- the theory is
  inconsistent in all but nine space and one time dimensions.
  Accordingly, this is also one of the great failures of string
  theory. The other theory is super-gravity, which is only consistent
  in eleven dimensions.
} 
We can therefore express any three-dimensional physical quantity or
law, such as Newton's equation of motion or Gauss' law, in any number
of integer dimensions $\nu \in \mathbb{Z}^+$. For example, we can
write down a corresponding quantity to (\ref{intthree}) in an
arbitrary number of dimensions,
\begin{eqnarray}
  Q(\nu;m)
  =
  \int_{{\mathbb R}^\nu} \! d^\nu k\, f_\nu({\bf k}; m) \ ,
\label{intnu}
\end{eqnarray}
where the wave vector ${\bf k}$ is now a $\nu$-dimensional vector, and
we integrate over the entire $\nu$-dimensional Euclidean space
${\mathbb R}^\nu$. I have placed a subscript on the integrand $f_\nu$
to indicate that it is determined by the theory expressed in $\nu$
dimensions. At this point, the spatial dimension is a non-negative
{\em integer}, so that $\nu=1,2,3,4, \cdots$. Since the integer $\nu$
is arbitrary in (\ref{intnu}), I have indicated that the corresponding
quantity contains $\nu$-dependence by writing $Q(\nu;m)$; however, for
notational simplicity I will drop the parametric dependence of
quantities such as mass and simply write $Q(\nu)$. A quantity that
diverges in three-dimensions will be finite in arbitrary $\nu$, but it
will typically exhibit a simple pole of the form
\begin{eqnarray}
  Q(\nu) =
  \frac{Q_0}{\nu-3} + Q_1(\nu) \ ,
\end{eqnarray}
where $Q_1(\nu)$ is finite at $\nu=3$.\footnote{\footnoteskip
  As we shall see in the next section, it is the nature of the Coulomb
  force that gives a pole in three dimensions, rather than some other
  value of the dimension. We shall further see that there is important
  physics in the residue $Q_0$ of the pole.
} 
Dimensional continuation is simply the act of treating $Q(\nu)$ as a
function of a complex argument $\nu$, {\em after} the integral
(\ref{intnu}) has been performed for {\em all} positive integer values
of $\nu$.  This is not an unfamiliar procedure, as the factorial
function $\nu!$ with $\nu \in {\mathbb Z}^+$ can be generalized to the
gamma function $\Gamma(\nu+1)$ in which $\nu \in {\mathbb C}$. Indeed,
given any function $Q(\nu)$ defined on the integers, with only mild
restrictions placed on the function at large values of the argument,
Carlson's Theorem~\cite{Carlsonth} allows us to continue this function
uniquely to the complex plane.

How does dimensional continuation work in practice? We will look at a
few specific examples in Sections~\ref{sec:physdim} and
\ref{sec:solution}, but for now let us consider a general physical
quantity $Q$ in which the integrand in (\ref{intthree}) depends solely
upon the modulus of ${\bf k}$, so that
\begin{eqnarray}
  Q = \int \! d^3 k\, f(k) \ . 
\label{Qthree}
\end{eqnarray}
In such a case it is not uncommon that the integrand in the
generalization (\ref{intnu}) is only a function of the modulus of the
$\nu$-dimensional wave vector ${\bf k}$, with the same functional form
as the integrand in (\ref{Qthree}). In other words, in (\ref{intnu})
we have $f_\nu({\bf k})=f(k)$ with $k=\vert {\bf k}\vert$, thereby
allowing us to write 
\begin{eqnarray}
  Q(\nu) = \int \! d^\nu k\, f(k) \ .
\end{eqnarray}
Since the integrand is a function only of $k$, we can extract the
angular integrals and write
\begin{eqnarray}
  Q(\nu) 
  = \!
  \underbrace{~~\int d\Omega_{\nu-1}~~}_{(\nu -\!1) \text{ integrals}}
  \cdot 
  \underbrace{\,\int_0^\infty k^{\nu-1} dk\, f(k)}_{
\text{~~~one-dimensional integral}} \ .
\label{Qfactor}
\end{eqnarray}
At this point, the dimension $\nu$ is simply an arbitrary positive
integer, $\nu \in {\mathbb Z}^+$. As we will show in the next
paragraph, the integration over all angles gives
\begin{eqnarray}
  \Omega_{\nu-1}
  \equiv 
  \int \!d\Omega_{\nu-1} = \frac{2 \pi^{\nu/2}}{\Gamma(\nu/2)} \ .
\label{intomega}
\end{eqnarray}
This leaves a one-dimensional integral to perform, in which 
$\nu$ simply acts as a parameter, 
\begin{eqnarray}
  F(\nu) \equiv \int_0^\infty dk\, k^{\nu-1} f(k) \ .
\label{intk}
\end{eqnarray}
The physical quantity now becomes the product of (\ref{intomega}) and
(\ref{intk}) with $\nu \in {\mathbb Z}^+$. In the~case of
(\ref{intomega}), we already know how to analytically continue $\nu$
to complex values. On the other hand, for (\ref{intk}) we can think
of $\nu$ as being an arbitrary complex number when performing the
one-dimensional integral over $k$, and therefore we can regard
$F(\nu)$ as a function over the complex $\nu$-plane, thereby giving
\begin{eqnarray}
  Q(\nu) 
  = 
  \frac{2 \pi^{\nu/2}}{\Gamma(\nu/2)}~ F(\nu) 
  ~~~~~\text{with}~~ \nu \in {\mathbb C} \ .
\end{eqnarray}
In this manner, we can regard $Q(\nu)$ as a function of a complex
argument $\nu$, and by Carlson's Theorem~\cite{Carlsonth}, this is the
unique continuation from positive integer values of $\nu$ to complex
values of $\nu$.

As an example of this procedure, let us prove (\ref{intomega}).
First, consider the one-dimensional Gaussian integral
\begin{eqnarray}
  \int_{-\infty}^\infty dk\, e^{-k^2} = \sqrt{\pi} \ .
\end{eqnarray}
If we multiply both sides together $\nu$ times (with
$\nu \in {\mathbb Z}^+$), we find 
\begin{eqnarray}
  (\sqrt{\pi}\,)^\nu 
  = 
  \int_{-\infty}^\infty \!\!dk_1\, e^{-k_1^2} 
  \int_{-\infty}^\infty \!\!dk_2\, e^{-k_2^2} 
  \, \cdots
  \int_{-\infty}^\infty \!\!dk_\nu\, e^{-k_\nu^2} 
  =
  \int \! d^\nu k\, e^{-{\bf k}^2} \ ,
\label{gaussnu}
\end{eqnarray}
where the wave vector ${\bf k}$ in the exponential of the last
expression is the $\nu$-dimensional vector ${\bf k}=(k_1, k_2, \cdots,
k_\nu)$, and ${\bf k}^2$ is the $\nu$-dimensional inner product ${\bf
k}^2=\sum_{\ell=1}^\nu k_\ell^2$. As in (\ref{Qfactor}), we can factor
the angular integrals out of the right-hand-side of (\ref{gaussnu}),
and the remaining one-dimensional integral can be converted to a Gamma
function with the change of variables $t=k^2$\,:
\begin{eqnarray}
  \pi^{\nu/2}
  = \!\!
  \int \!\! d\Omega_{\nu-1} \cdot \!\!
  \int_0^\infty \!\!\! dk\, k^{\nu-1}\,e^{-k^2} 
  \! = \!\!
  \int \!\! d\Omega_{\nu-1} \cdot 
  \frac{1}{2}\int_0^\infty \!\! dt\, t^{\nu/2-1}\, e^{-t} 
  = \!\!
  \int \!\! d\Omega_{\nu-1} \cdot \frac{1}{2}\, \Gamma(\nu/2) \,.
\label{pinuomega}
\end{eqnarray}
Solving for $\int\!d\Omega_{\nu-1}$ in (\ref{pinuomega}) gives
(\ref{intomega}).  As an aside, it is interesting to note that we have
also found the hyper-area of a \hbox{$(\nu\! -\! 1)$-dimensional}
sphere of radius $r$ in ${\mathbb R}^\nu$,
\begin{eqnarray}
  A_\nu = \frac{2\,\pi^{\nu/2}}{\Gamma(\nu/2)}~ r^{\nu-1} \ ,
\label{areanu}
\end{eqnarray}
and integrating (\ref{areanu}) gives the hyper-volume of a
$\nu$-dimensional ball of radius $r$,
\begin{eqnarray}
  \Sigma_\nu = 
  \frac{\pi^{\nu/2}}{\Gamma(\nu/2+1)}~ r^\nu \ .
\end{eqnarray}

\pagebreak
\subsection{Physics of Dimensional Continuation}
\label{sec:physdim}

So far we have only introduced dimensional continuation as a
regularization prescription, a means by which infinite theories can be
rendered temporarily finite. This is the use to which dimensional
continuation is commonly employed in field theory; however, this alone
is not sufficient to render the method suitable for our
purposes. Instead, there are also physics reasons that make dimensional
continuation applicable to the problem at hand. I will now show that
the Coulomb potential, and indeed physics in general, behaves
differently in different spatial dimensions. In particular,
short-distance or ultraviolet (UV) physics dominates when $\nu>3$;
while long-distance or infrared (IR) physics dominates when
$\nu<3$. In $\nu=3$ dimensions, both UV and IR physics are equally
important.\footnote{\footnoteskip
  There are a number of such {\em coincidences}, both physical and
  mathematical, that suggest there is something special about living
  in three spatial dimensions and one time dimension. This could well
  be the Anthropic Principle at work. 
} 
This means that dimensional continuation acts as a ``physics sieve,''
allowing us to capture the leading UV and IR physics simply by
performing the relevant integrals in dimensions greater than or less
than the traditional $\nu=3$. In the next few paragraphs, we will
discuss why changing the dimension of space emphasizes either
long- or short-distant physics. Understanding this point is 
crucial for all that follows. 

We now turn to finding the $\nu$-dimensional Coulomb potential, an
exercise that succinctly illustrates how the physics of a system
changes with the dimension of space. Besides illustrating that UV
physics dominates in higher dimensions (and conversely), this example
will also clarify the manner by which one performs physical
calculations in arbitrary dimensions. Let us consider Poisson's
equation
\begin{eqnarray}
  {\bm\nabla}\!\cdot\!{\bf E}({\bf x}) 
  = \rho({\bf x}) \ .
\label{peq}
\end{eqnarray}
There is nothing in this equation {\em per se} that restricts us to
three dimensions.\footnote{\footnoteskip 
  One can most easily write the complete set of Maxwell's equation in
  a general dimension by employing the Lorentz covariant form
  ${\sum}_{\alpha =0}^\nu \partial F^{\alpha \beta}/\partial x^\alpha
  = j^\beta$, where the electric and magnetic fields have been
  expressed in terms of the anti-symmetric field tensor
  $F^{\alpha\beta}$.
} We {\em choose} to describe the electric field ${\bf E}$ and the
spatial coordinate ${\bf x}$ as three-dimensional vectors because we
live in three dimensions. However, from the mathematics alone, we
could equally well consider these vectors as living in an arbitrary
$\nu$-dimensional space ${\mathbb R}^\nu$. The rectilinear coordinates
would then become ${\bf x}= (x_1, x_2, \cdots, x_\nu)$, with a similar
expression for the \hbox{$\nu$-dimensional} electric field, while the
gradient would be ${\bm\nabla}=\left( \partial/\partial x_1,
\partial/\partial x_2, \cdots, \partial/\partial x_\nu\, \right)$. It
is convenient to write Poisson's equation in its integral
representation, which will allow us to calculate the electric field of
a point charge with relative ease. Consider a point-charge at the
origin given by $\rho({\bf x})= e\,
\delta^{\scriptscriptstyle(\nu)}({\bf x})$, and let $\Sigma$ be any
volume containing the charge $e$. Then, in a general number of
dimensions, we can integrate (\ref{peq}) to obtain
\begin{eqnarray}
  \int_\Sigma d^\nu x \,{\bm\nabla}\!\cdot\!{\bf E} = e \ .
\end{eqnarray}
Note that the dimensionality of space is now explicitly indicated by
the integration measure.

For our purposes, the advantage of the integral representation of
Poisson's equation is that the electric field of a point charge in the
$\nu$-dimensional space can easily be calculated by the same symmetry
principles that hold in three dimensions. Suppose the volume $\Sigma$
is a spherical ball $B_r$ of radius $r$ centered on the charge. The
boundary of $B_r$ is a sphere of dimension $\nu\!-\!1$ and will be
denoted by $\partial B_r$ (I am using the common notation $\partial$
in differential geometry for the boundary of a manifold). From
rotational symmetry, the electric field ${\bf E}$ of a point charge is
directed radially outward and lies normal to the surface $\partial
B_r$ at each point. We will denote the magnitude of the electric field
at radius $r$ by $E(r)$.  Recall that in (\ref{areanu}), we calculated
the hyper-area of the $(\nu\!-\!1)$-sphere $\partial B_r$ to be
$A_\nu\, = \Omega_{\nu-1} r^{\nu-1}$ with $\Omega_{\nu-1}=\pi^{\nu/2}/
\Gamma(\nu/2)$. Since the divergence theorem holds in an arbitrary
Euclidean space (like the laws of physics, there is nothing in the
divergence theorem to restrict the dimensionality of space to three),
we can write
\begin{eqnarray}
  e 
  = 
  \int_{B_r} \!\! d^\nu x \,{\bm\nabla}\!\cdot\!{\bf E}
  = \oint_{\partial B_r}\!\! d{\bf A}\!\cdot\! {\bf E}
  =
  \Omega_{\nu-1}\, r^{\nu-1} \cdot  E(r) \ .
\end{eqnarray}
At position ${\bf x}$, the electric field therefore takes the form
\begin{eqnarray}
  {\bf E}({\bf x}) = \frac{e}{ \Omega_{\nu-1}\, r^{\nu-1}}\,
  \hat{\bf x} \ ,
\label{E}
\end{eqnarray}
where I am using the notation ${\bf x}= r\, \hat{\bf x}$, with
$\hat{\bf x}$ being a unit vector pointing in the direction of ${\bf
x}$ and $r=\vert {\bf x} \vert$ being the magnitude.  It is often more
convenient to work with the electric potential, a scalar quantity
$\phi_\nu(r)$ defined by $E(r) = - d\phi_\nu(r)/dr$. In fact, we need
the potential energy \hbox{$V_\nu(r) = e\, \phi_\nu(r)$}, so that
\begin{eqnarray}
  V_\nu(r)
  = 
  \frac{1}{C_\nu}\,  \frac{e^2}{r^{\nu-2}} \ ,
\label{Vnu}
\end{eqnarray}
with 
\begin{eqnarray}
  C_\nu 
  = 
  \Omega_{\nu-1}\,(\nu-2) 
  = 
  \frac{4 \pi^{\nu/2}}{\Gamma(\nu/2-1)} \ .
\end{eqnarray}
For $\nu\!=\!3$ we have $C_3=1/4\pi$, which is the origin of the $4
\pi$ of rationalized units. Note from (\ref{Vnu}) that the engineering
unit of electric charge is a function of the dimension $\nu$. This is
because the $\nu$-dimensional Coulomb potential (\ref{Vnu}) must have
units of energy, and consequently the engineering unit of $e^2$ is
$\text{Energy} \times (\text{Length} )^{\nu-2}=\text{Mass} \times
(\text{Length})^\nu/(\text{Time})^2$.  It is quite natural that a
composite quantity, made from the fundamental units of Mass, Length
and Time, change its engineering dimension with the dimension of
space.

\begin{figure}
\includegraphics[scale=0.45]{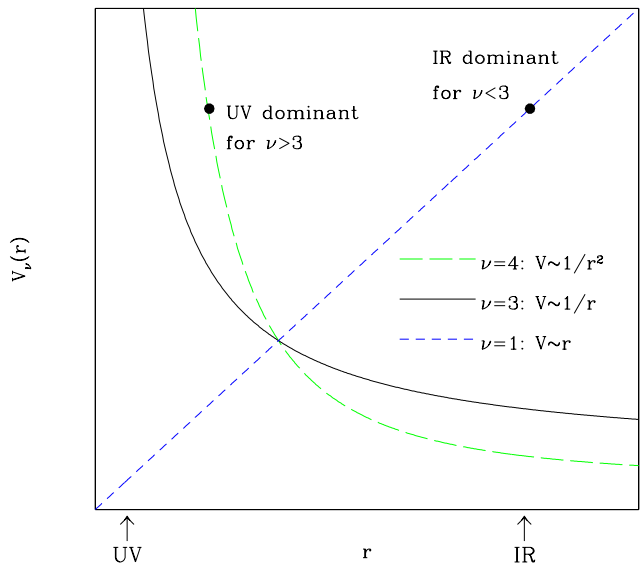}
\vskip-0.8cm 
\caption{\captionskip
Short-distance or ultraviolet (UV) physics dominates in dimensions
$\nu>3$. Long-distance or infrared (IR) physics dominates when
$\nu<3$. UV and IR physics are equally important in $\nu=3$.  
}
\label{fig:coulomb}
\end{figure}

Figure~\ref{fig:coulomb} shows the Coulomb potential for $\nu\!=\!3$,
along with two representative dimensions on either side of
$\nu\!=\!3$. As the figure illustrates, the short-distance or UV
behavior of the Coulomb potential becomes more severe as the dimension
increases above $\nu\!=\!3$, while the long-distance or IR behavior
dominates for dimensions below $\nu=3$. The arbitrary integration
constant for the potential energy has been adjusted in each case so
that all three graphs intersect at a single point. This was purely for
aesthetics, as it renders the differences between the potentials more
apparent. Despite the trouble we went through in the previous
paragraph to find the coefficients $C_\nu$, in this paragraph (and
only in this paragraph) I have temporarily set $C_\nu\!=\!1$. This
will make it easier to compare the $r$-dependence of various
potentials, and I would rather opt for clarity over notational
consistency.  For the representative potential with dimension below
$\nu=3$, I chose to graph the one-dimensional potential
\hbox{$V_1(r)=e^2 r$} rather than the two-dimensional potential
\hbox{$V_2(r)=e^2\ln(r/r_0)$}, where $r_0$ is an arbitrary integration
constant with units of length. In both cases the potential increases
without bound at long distances, thereby illustrating the dominance of
long-distance or IR physics in dimensions less than three.  However, I
chose to graph $V_1(r)$ rather than $V_2(r)$ because the latter
possesses a potentially misleading divergence as $r \to 0$: unlike the
short-distance or UV divergence associated with the potential in
dimensions greater than three, the $r\to 0$ divergence of $V_2(r)$ is
integrable, and consequently harmless.  Therefore, for purposes of
illustration, the potential $V_1(r)= e^2\, r$ makes the point better
than $V_2(r)= e^2\,\ln(r/r_0)$.  Remarkably, we now see that by simply
selecting the dimension $\nu$, we can dial a potential $V_\nu(r)$ that
filters either long-distance or short-distance physics.

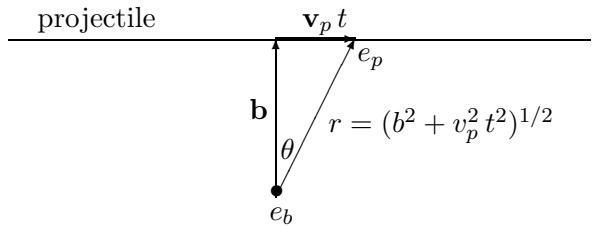
\begin{figure}[t]
\begin{picture}(120,120)(-75,0)

\put(-101,60){\line(1,0){220}}

\put(0,0.5){\vector(0,1){60}}
\put(-2.3,0){$\bullet$}
\put(-2.3,-8){$e_b$}
\put(-10,30){${\bf b}$}

\put(0,60.5){\vector(1,0){30}}
\put(10,65){${\bf v}_p\, t$}
\put(-90,65){projectile}

\put(0,0.5){\vector(1,2){30}}
\put(31,52){$e_p$}
\put(1.9,15){$\theta$}
\put(20,25){$r=(b^2 + v_p^2\, t^2)^{1/2}$}

\end{picture}
\caption{\captionskip 
  A projectile of charge $e_p$, mass $m_p$, and velocity ${\bf v}_p$
  passing a fixed charge $e_b$.  The impact parameter ${\bf b}$ is
  normal to the velocity, so that ${\bf b}\cdot{\bf v}_p=0$. The
  radial separation between the charges is \hbox{$r(t)=(b^2 + v_p^2\,
  t^2)^{1/2}$}, and only the $b$-component $E_\smPerp = E\cos\theta =
  (b\, e_b/\Omega_{\nu-1})\, r^{-\nu}$ of the electric field
  contributes the impulse integration (\ref{delp}), where $\cos\theta=
  b/r$.
}
\label{fig:scat}
\end{figure}

I would now like to show how the pole at $\nu=3$ arises from the
$\nu$-dimensional Coulomb potential. In the temperature equilibration
process, individual plasma species exchange energy by mutual Coulomb
interactions. For example, consider a particle in the plasma with
charge $e_p$, mass $m_p$, and velocity ${\bf v}_p$, and suppose it
passes another charge $e_b$ with an impact parameter ${\bf b}$. This
is illustrated in Fig.~\ref{fig:scat}, where, to zeroth order, the
projectile follows the straight line ${\bf x}(t)={\bf b} + {\bf v}_p
\, t $ as a function of the time $t$, with ${\bf b}\cdot {\bf
v}_p=0$.\footnote{\footnoteskip
  In this simple example, I am not concerned with hyperbolic orbit
  corrections and the like; but rather, I am tracing the origin of the
  logarithmic divergence of the Coulomb potential and the pole at
  $\nu=3$. For this, we can work with a hot dilute plasma where a
  linear trajectory will suffice.
} In $\nu$-dimensional space, the Coulomb potential is given by
(\ref{Vnu}), and the corresponding electric field ${\bf E}$ by
(\ref{E}), with $e$ replaced by $e_b$. The projectile therefore
acquires a momentum transfer
\begin{eqnarray}
  \Delta {\bf p} 
  &=& 
  e_p \int_{-\infty}^{+\infty} \!\! dt\,
  {\bf E}({\bf b} + {\bf v}_p\, t) \ ,
\label{delP}
\end{eqnarray}
and it suffers a corresponding change in energy
\begin{eqnarray}
  \Delta E
  &=& 
  \frac{{\Delta{\bf p}}^2}{2m_p}  \ .
\label{delE}
\end{eqnarray}
The component of the electric field along the direction of motion
${\bf v}_p$ integrates to zero in (\ref{delP}), while the component
normal to the trajectory gives the impulse
\begin{eqnarray}
  \Delta {\bf p} 
  &=& 
  e_p \int_{-\infty}^{+\infty} \!\! dt\,E_\smPerp({\bf b} + 
  {\bf v}_p\, t) \, \hat{\bf b}
  =
  \frac{e_p\, e_b}{\Omega_{\nu-1}}
  \int_{-\infty}^{+\infty} \!\! dt\, 
  b\, (b^2 + v_p^2\, t^2)^{-\nu/2}\, \hat{\bf b}
\\[5pt]
    &\sim&
  \frac{1}{b^{\nu-2}}\, \hat{\bf b}   \ .
\label{delp}
\end{eqnarray}
The temperature equilibration rate contains a factor involving the
cross-section weighted energy transfer, and we see that the energy
exchange between $p$ and $a$ can be written
\begin{equation}
  \frac{dE}{dt}
  \sim
  \int \!d\sigma\, \Delta E
  \sim
  \int_{b_\text{min}}^{b_\text{max}} \, \frac{db}{b^{\nu - 2} } \, \ ,
\label{dedtsim}
\end{equation}
where we have used the fact that $d\sigma\cdot\Delta E \sim
b^{\nu-2}db \,\cdot\, b^{-2(\nu-2)} \sim db/b^{\nu-2}$. I will elaborate
further on this example in the next lecture, but for now one should
simply note that the rate (\ref{dedtsim}) implies that large $\nu$ is
dominated by short-distance physics and small $\nu$ is dominated by
long-distance physics.  Moreover, expression (\ref{dedtsim}) gives
$\nu = 3$ as the dividing line between these two regions. To see this,
note that for $\nu > 3$ the impact parameter integral is not sensitive
to the large distance cut off, and we may simply take the limit 
${b_\text{max}} \to \infty$ to obtain
\begin{equation}
  \nu > 3 \,: \hskip1cm 
  \int_{b_\text{min}}^\infty \frac{db}{b^{\nu-2}} 
  = 
  \frac{b_\text{min}^{3-\nu}}{\nu -3} 
  ~~~\Rightarrow~~
  \text{UV dominant and pole at}~\nu=3 \ .
\end{equation}
Conversely, for $\nu < 3$, we can set $ {b_\text{min}} = 0 $, with
\begin{equation}
  \nu < 3 \,: \hskip1cm 
  \int_0^{b_\text{max}} \frac{db}{b^{\nu-2}} 
  = 
  \frac{b_{\text{max}}^{3-\nu}}{3-\nu} 
  ~~~\Rightarrow~~
  \text{IR dominant and pole at}~\nu=3 \ .
\end{equation}
The results displayed are the dominant forms in the two regions of
spatial dimensionality about $\nu=3$. In either case, the stopping
power contains a pole $1/(\nu-3)$.

There are a number of important consequences arising from the UV and
IR behaviors of the Coulomb potential (\ref{Vnu}). I will discuss this
more fully in Section~\ref{sec:dimred}, but for now recall that the
derivation of the Boltzmann equation, as presented in
Ref.~\cite{huang} for example, breaks down for the Coulomb potential
in three spatial dimensions because of the aforementioned infrared
singularity. However, in $\nu>3$ the ``textbook derivation'' of the
Boltzmann equation with the Coulomb potential (\ref{Vnu}) is finite
and completely rigorous. The simple pole at $\nu=3$ in the scattering
kernel corresponds to an IR divergence because of the long-range
nature of the Coulomb force in three dimensions.  Furthermore, because
dimensions greater than three enhance the UV physics, the classical
Born, Bogoliubov, Green, Kirkwood, and Yvon (BBGKY) hierarchy reduces
to the Boltzmann equation to leading order in~$g$ (or to leading order
in the number density) when $\nu>3$. A similar reduction from BBGKY
holds for the Lenard-Balescu equation in $\nu < 3$, and the ``textbook
derivation'' \cite{lifs} is also rigorous in these dimensions. In
$\nu=3$, the derivations of the Boltzmann and Lenard-Balescu equations
break down for the Coulomb potential.  This is not because the $\nu=3$
version of the BBGKY hierarchy is divergent, but because the
truncation procedure that leads to the Boltzmann and Lenard-Balescu
equations breaks down for the three dimensional Coulomb
potential.\footnote{\footnoteskip
  For example, in deriving the Boltzmann equation from BBGKY, one
  invokes a principle of uncorrelated collisions to replace the
  two-point correlation function by the product of two one-point
  functions, and this leads to the infrared divergence. This means, of
  course, that it is the long-distance correlations themselves that
  tame the IR divergence of the Boltzmann equation, correlations that
  are neglected by the truncation process. As a related point, this
  procedure imposes an implicit initial condition, thereby providing a
  {\em direction of time} for the Boltzmann equation (and this occurs
  even for short-range potentials).
}
Indeed, BBGKY is completely finite for the Coulomb potential in
$\nu=3$, albeit completely useless for our purposes. Section~II of
BPS~\cite{bps} provides more details, especially the two paragraphs
between Eqs.~(2.7) and (2.8).

\section{The Lamb Shift and the Coulomb Logarithm}
\label{sec:lamb}

The Lamb shift is interesting for us because it provides another
connection with the Coulomb logarithm, both in the historical details
and in much of the physics. The Lamb shift is a small energy split in
the otherwise degenerate $2S$ and $2P$ states of total angular
momentum $j=1/2$. The measured value is about $\Delta E_\text{lamb}
\simeq 4.4 \times 10^{-6}\, {\rm eV}$, with the $2S_{1/2}$ state lying
above the $2P_{1/2}$ state. Therefore, when an electron makes a
transition from the $S$-state to the $P$-state, it emits a microwave
photon of frequency $\Delta \nu_\text{\,lamb} \simeq 1060\,{\rm MHz}$.
Calculating the observed value of the Lamb shift was the first great
success of quantum electrodynamics (QED), the relativistic quantum
theory of light and matter.\footnote{\footnoteskip
  The other early success of QED, which followed soon after the Lamb
  shift, was calculating the magnetic dipole moment of the
  electron. It is customary to write the magnetic moment of the
  electron $\mu_e$ in terms of the Bohr magneton $\mu_\smB= e\hbar/2
  m_e$ by introducing the dimensionless $g$-factor: $\mu_e = g_e\,
  \mu_\smB$. Using relativistic single-particle quantum mechanics, in
  1928 Dirac predicted $g_e=2$ (exactly). In 1948 Schwinger used QED
  to calculate the radiative corrections, and he found $g_e/2= 1+
  \alpha/2\pi + {\cal O}(\alpha^2) = 1.0011614$, which was in
  excellent agreement with experiment. Today, the electron's magnetic
  dipole moment has been calculated to include $\alpha^4$ terms, and
  is in agreement with experiment to 10 significant figures, the most
  accurately verified quantity in the history of physics.
%
}

Dirac's relativistic model of the hydrogen atom, {\em i.e.} his
relativistic theory of the electron in a Coulomb potential, made the
prediction that the energy levels of the hydrogen atom (neglecting the
hyperfine structure) depend only upon the principal quantum number $n$
and the total angular momentum $j$ (the sum of the orbital momentum
$\ell$ and the spin $s=1/2$ of the electron). In particular, Dirac
predicted that the $2S_{1/2}$ and $2P_{1/2}$ states should be
degenerate. At the 1947 Shelter Island conference in New York, W.~Lamb
and R.~Retherford announced the results of their highly sensitive
experiment measuring the emission frequency of photons in a $2S$-$2P$
transition, thereby establishing the unequivocal experimental
existence of the Lamb shift.\footnote{\footnoteskip
  In the 1930s, S.~Pasternak analyzed experimental data suggesting
  that such an energy split might exist; however, the systematic error
  was as large as the energy splitting itself.
}
While almost degenerate compared to the binding energy of the hydrogen
atom, today's accepted experimental splitting is~\cite{lowelqft}
\begin{eqnarray}
  \Delta E_\text{\,lamb} 
  &\equiv&
  E_{2S_{1/2}} - E_{2P_{1/2}}
  = 4.374898(7) \times 10^{-6}\,{\rm eV} 
\\[5pt]
  \Delta \nu_\text{\,lamb} 
  &\equiv&
  \frac{\Delta E_\text{lamb}}{h}
  = 
  1057.845(9)\,{\rm MHz} \ ,
\end{eqnarray}
where Plank's constant in the form $h=4.135\,667\,33(10) \times
10^{-15}\, \text{eV-s}$ is the conversion factor between energy and
frequency. Hans Bethe had attended the Shelter Island conference, and
on the train ride back he performed the first calculation of the Lamb
shift, finding the value \hbox{$\Delta \nu_\text{\,bethe} = 1040\,{\rm
MHz}$}.  Bethe's calculation was recognized as being only a rough
approximation, neglecting high energy relativistic effects, but its
close agreement with experiment was cause for optimism.

The dominant contribution to the Lamb shift comes from the bound $2S$
and $2P$ electrons exchanging virtual photons with the atomic
nucleus.\footnote{\footnoteskip \label{foot:othercorrections}
  There are also several other subdominant mechanisms, such as the
  vacuum polarization of the photon and the anomalous magnetic dipole
  moment of the electron. These give, respectively, the contributions
  $\Delta \nu_\text{lamb}^\text{vac}=- 27\,{\rm MHz}$ and $\Delta
  \nu_\text{lamb}^\text{mag}=+68\,{\rm MHz}$.
} These radiative corrections effectively smear the point-like nature
of the nucleus, thereby altering the atomic energy levels (the
electron no longer ``sees'' a pure Coulomb potential). Any possible
number of photon exchanges with the nucleus are permitted, from a
single high-energy photon (hard/UV physics) to many low-energy photons
(soft/IR physics), and this means there are two disparate but
competing energy scales in the problem.  In a manner similar to the
Coulomb energy-loss rate (\ref{ratemodel}), the Lamb shift takes the
form
\begin{eqnarray}
  \Delta E_\text{\,lamb} 
  =
  K\, \ln\!\left\{\frac{E_\text{max}}{E_\text{min}}\right\} \ .
\label{lambscematic}
\end{eqnarray}
The UV scale is set by $E_\text{max} \sim m_e c^2 = 511\,{\rm keV}$
and the IR scale $E_\text{min}$ is set by the binding energy of the
hydrogen atom
\begin{eqnarray}
  E_0 
  =
  \frac{1}{2}\,\left(\frac{e^2}{4\pi \hbar}\right)^2 m_e
  =
  \frac{1}{2}\, \alpha^2 \, m_e\, c^2 = 13.6 \, {\rm eV} \ .
\label{E0}
\end{eqnarray}
To understand the origin of these scales, note that $E_\text{max}$
takes its value from the energy at which relativistic effects for the
electron become important (the rest-mass energy of the electron),
while $E_\text{min}$ takes its value from the only low-energy scale in
the problem (namely, the binding energy of the atom). The soft photon
IR interactions can be handled by nonrelativistic means, while the
hard UV photons require a more complicated relativistic treatment.
For this reason, Bethe used the simpler nonrelativistic formalism,
cutting his calculation off at the relativistic energy scale
$E_\text{max} \sim m_e c^2$ at which his formalism broke down. This is
akin to cutting the calculation of the energy exchange rate off at
some small distance scale $b_\text{min}$.  Bethe then concentrated on
the low energy theory, which is analogous to looking at the
Lenard-Balescu equation in the rate calculation. Bethe found a
logarithmic UV divergence, in the same way the Lenard-Balescu equation
has a UV divergence, but he was able to regularize the infinity by
applying a technique known as mass renormalization, thereby rendering
his calculation finite. His calculation was completely rigorous at low
energies, and Bethe found the coefficient $K$ and the low energy
cutoff $E_\text{min}$, both {\em exactly}~\cite{sak}:
\begin{eqnarray}
  K = \frac{\alpha^3}{3\pi}\,E_0
  ~~~\text{and}~~~ 
  E_\text{min} = A_0\, E_0 
  ~~~\text{with}~~~ E_\text{max} \sim m_e\, c^2 \ .
\label{KBethe}
\end{eqnarray}
The binding energy $E_0$ of the hydrogen atom is given by (\ref{E0}),
and while I will not write it down, the coefficient $A_0$ is
rigorously defined in terms of sums of matrix elements of the various
intermediate states. These matrix elements are sufficiently
complicated that one can only calculate them numerically, and to three
significant figures Bethe found $A_0 = 17.8$. Bethe's exact
calculation of $E_\text{min}$ is akin to an exact calculation of
$b_\text{max}$ in the rate problem.  This, however, is a point where
the analogy is not precise: Bethe was able to exactly calculate
$E_\text{min}$ through the regularization procedure of mass
renormalization, which has no counterpart in plasma physics. On the
other hand, Lyman Spitzer was only able to estimate the value of the
maximum impact parameter in the Coulomb logarithm to be of order
$b_\text{max} \sim \kappa_\smD^{-1}$. Bethe was able to perform this
feat because QED is a fundamental theory of nature, while the
Boltzmann equation is not.  Using (\ref{E0}), we can express
(\ref{lambscematic}) as
\begin{eqnarray}
  \Delta E_\text{\,lamb}^\smB
  = 
  -\frac{\alpha^5}{3\pi}\,m_e\,c^2 \,
  \ln\{C_\smB \cdot \alpha \} 
  ~~~\text{with}~~
  C_\smB = \sqrt{17.8/2} = 2.98 \  ,
\label{DelEBethe}
\end{eqnarray}
where $\alpha=e^2/4\pi \hbar c = 1/137.036$ is the fine structure
constant. Bethe's calculation therefore gives $\Delta
E_\text{lamb}^\smB = 4.3 \times 10^{-6}\,{\rm eV}$ or $\Delta
\nu_\text{lamb}^\smB=1040\,{\rm MHz}$. 

Bethe did not, however, calculate the {\em exact} coefficient under
the logarithm.  Since his calculation broke down at relativistic
energies, he used the somewhat {\em ad hoc} value \hbox{ $E_\text{max}
= m_e\, c^2$} (rather than some multiple thereof) for the UV
cutoff. It turns out that the high energy corrections to
(\ref{DelEBethe}) are rather small, so Bethe's result was perhaps more
accurate than warranted.  Since Bethe only estimated the maximum
energy cutoff, his result (\ref{DelEBethe}) is only accurate to
leading order in $\alpha$, and the constant under the logarithm
required a more accurate treatment. The analogy further continues:
Bethe calculated the Lamb shift to leading order, just as
Spitzer calculated the Coulomb logarithm to leading order. Both
calculations found the correct coefficient in front of the logarithm,
and the order of magnitude of the argument of the logarithm. These
calculations failed to extract the exact constant under the logarithm,
although Bethe managed to find the exact expression for the low-energy
cutoff. In this sense, Bethe is the Spitzer of the Lamb shift.

Shortly after this, R.~Feynman and J.~Schwinger independently
calculated the high energy contribution using their respective
relativistically covariant formalisms,\footnote{\footnoteskip
  F.~Dyson soon proved that Schwinger's mathematically rigorous
  formalism was equivalent to Feynman's more intuitive but easier to
  use approach. Along with the Japanese physicist S.~Tomonaga,
  Schwinger and Feynman shared the 1965 Nobel prize ``for their
  fundamental work in quantum electrodynamics, with deep-ploughing
  consequences for the physics of elementary particles.''
} and their calculations were exact to leading and next-to-leading
order in $\alpha$. After adding their contribution of the high energy
corrections to Bethe's low energy form, the calculation of the Lamb
shift was complete and the constant under the logarithm was fully
determined.  Or so it would seem. Simultaneously, J.~French and
V.~Weisskopf completed an independent calculation of the Lamb shift
that disagreed with Feynman-Schwinger, albeit only slightly in the
coefficient under the logarithm. The plot thickened as yet another
calculation of the Lamb shift was completed that agreed with
French-Weisskopf, this time from N.~Kroll and Lamb
himself.\footnote{\footnoteskip
 Lamb was both an experimentalist and a theorist, and he received the
 1955 Nobel Prize for his ``discoveries concerning the fine structure
 of the hydrogen spectrum.''
} So now we seem to have a real problem: the Schwinger-Feynman
calculations agree with each other, but disagree with the
French-Weisskopf and Kroll-Lamb calculations. Schwinger and Feynman
had developed independent but equivalent formalisms that manifestly
exhibited the relativistic covariance of the theory, while the other
four had used a cumbersome formalism developed in the 1930's (now
called {\em old-fashioned perturbation theory}).  As it turns out,
Schwinger and Feynman had made the same subtle mistake, and
Weisskopf's calculation was correct. The contribution to the Lamb
shift we have been considering has the form
\begin{eqnarray}
  \Delta E_\text{lamb} = \frac{\alpha^3}{3\pi}\, E_0 \,
  \left( \ln\!\left\{ \frac{m_e\, c^2}{2 E_\smB}\right\} 
  + \frac{91}{120} \right) \ ,
\label{lambCorrect}
\end{eqnarray}
where $E\smB = 17.8 E_0$ is Bethe's low energy result (\ref{KBethe}).
As with the BPS Coulomb logarithm, we can bring the additive constant
in (\ref{lambCorrect}) inside the logarithm, and using (\ref{E0}) we
can express (\ref{lambCorrect}) as
\begin{eqnarray}
  \Delta E_\text{lamb} 
  = 
  -\frac{\alpha^5}{3\pi}\, m_e c^2 \, \ln\{C\,\alpha \}
  ~~~\text{with} ~~
  C = 2.98 \ .
\label{explambCorrect}
\end{eqnarray}
In terms of the Bethe's coefficient $C_\smB$ of (\ref{DelEBethe}), the
complete Lamb shift coefficient is \hbox{$C = \sqrt{2}\, e^{-91/240}
\, C_\smB$} (apart from the photon vacuum polarization and the
electron dipole moment corrections, which were mentioned in
footnote~\ref{foot:othercorrections}, but will not concern us
further).

French spent the next year tracing down the origins of the discrepancy
between the calculations. He could not find an error in the
Feynman-Schwinger high energy calculation, nor was there an error in
Bethe's low energy calculation.  Instead, the error was in the way the
high-energy calculation of Feynman-Schwinger was ``joined'' onto the
low-energy result of Bethe. The high-energy calculation is also
infinite and in need of regularization, but this time it contains an
IR divergence. This is analogous to the Boltzmann equation containing
the correct short-distance physics, but nonetheless diverging at long
distances. Therefore, Feynman-Schwinger had to introduce an
intermediate step in which they regulated their high energy theories
in the infrared. They chose to do this by assigning a small mass to
the photon, and then taking this mass to zero at the end of the
calculation. While this is a common IR regularization scheme in QED,
it is incommensurate with Bethe's regularization scheme in his low
energy calculation, where he used a simple cutoff procedure in a
high-energy integral. French-Weisskopf and Kroll-Lamb had gotten the
correct result because they calculated both the low-energy and the
high-energy contributions using the {\em same} formalism, and
therefore with consistent regularization schemes for both short and
long distances.  For more on the consequences of regulating large and
small scales in an incommensurate manner, see Feynamn's footnote~13 in
Ref.~\cite{thirteen} of this lecture, which I have quoted in the
bibliography in its entirety. That one must calculate the large and
small scales in exactly the same manner is not a minor point, as
underscored by the stature of the physicists who failed to realize its
importance when they first calculated the Lamb shift from the then
fledgling theory of quantum electrodynamics (QED). 

There are a number of consistent regularization schemes in use today
in QED and other field theories of nature, with dimensional
regularization being one of the most popular and easy to use. These
regularization and renormalization schemes have allowed us to
calculate a great many experimentally verified quantities, to
extremely high precision, and there is no longer any doubt in their
correctness. While extending the dimension of space to complex values
might at first seem unsettling, this procedure works. When dimensional
continuation was first introduced into quantum field theory, there
were strong reactions against it. However, calculations performed with
this method agree with calculations using other regularization
schemes, and more important, with experiment. In time, particle
physicists learned to accept the notion that one can perform correct
three dimensional calculations by working in arbitrary complex
dimensions. For our concerns, we note that Refs.~\cite{lfirst} and
\cite{lowelqft} have indeed calculated the Lamb shift using the method
of dimensional continuation. The problem of Feynman-Schwinger is
avoided, and dimensional continuation gives the correct experimentally
observed result with much less effort than more traditional
methods. As the above discussion illustrates, this is no small
achievement.

In summary, dimensional continuation is powerful because (i) it
is a consistent regularization scheme that (ii) lends itself to a
perturbative analysis, and (iii) requires relatively simple (or at
least easily learned) calculational tools. 
DeWitt's calculation~\cite{gould} was certainly consistent, in that it
matched the long and short-distance physics commensurately [to order
${\cal O}(g^3)$]. This is because he starts with a finite regulated
theory that treats long- and short-physics together within a single
framework, albeit with a regularization scheme that does not lend
itself to a systematic perturbative analysis. By comparison with the
BPS result, we know DeWitt was accurate to order $g^2$,
inclusive. But, as illustrated by Feynman's footnote 13 of
Ref.~\cite{thirteen}, and by the story of the Lamb shift above, any
attempt at treating the long- and short-distance physics by separate
regularization schemes (as much of the plasma literature currently
does), will likely miss the very constants they are trying to
calculate. Recall, it took two correct calculations to find the error
in Feynman's single calculation. Dimensional continuation performs all
necessary book keeping, at both small and large scales, and in a manner
that affords simple calculations and perturbative expansions.

\section{Calculating the Rate Systematically with Dimensional Continuation}
\label{sec:solution}

\subsection{Dimensional Reduction of BBGKY}
\label{sec:dimred}

Let us return to the rate equation (\ref{dedtbe}) in the light of the
apparatus of dimensional continuation that we have constructed.  Since
we are interested in spatially uniform plasmas, we will only consider
particle distributions that are functions of the momentum, or
equivalently the velocity.  Let ${\bf v}_\nu$ denote a
\hbox{$\nu$-dimensional} velocity vector with components $v_\ell$
for~$\ell=1, \cdots, \nu$, and define a $\nu$-dimensional distribution
function $f_\nu({\bf v}_\nu,t)$ by
\begin{eqnarray}
  d^\nu v \,f_\nu({\bf v}_\nu,t) \equiv \text{number of particles in a
  hyper-volume $d^\nu v$ about ${\bf v}_\nu$ at time $t$} \ .
\label{fnu}
\end{eqnarray}
Then the generalization of the three dimensional result (\ref{dedtbe})
to $\nu$-dimensions would be
\begin{eqnarray}
  \frac{d {\cal E}_\nu}{dt} 
  = 
  \int d^\nu v \,\frac{1}{2}\, m\, v_\nu^2 ~
  \frac{\partial f_\nu}{\partial t}({\bf v}_\nu,t)\ ,
\label{dedtbenu}
\end{eqnarray}
where the square of the velocity in (\ref{dedtbenu}) is $v_\nu^2 =
{\bf v}_\nu \cdot {\bf v}_\nu = \sum_{\ell=1}^\nu v_\ell^2$.  As
previously mentioned, the standard textbook calculation of the
Boltzmann equation goes through without an infrared divergent
scattering kernel when $\nu>3$. For now, I will write this equation in
schematic form as
\begin{eqnarray}
  \frac{\partial f_\nu}{\partial t} + 
  {\bf v}_\nu \! \cdot \!{\bm\nabla}  f_\nu 
  &=& 
  B_\nu[f]   ~~~~:\, \nu> 3 \ .
\label{BEsimpnu}
\end{eqnarray}
Note that ${\bm\nabla}$ is the $\nu$-dimensional gradient and ${\bf
v}_\nu \! \cdot \!{\bm\nabla} f_\nu= \sum_{\ell=1}^\nu
v_{\ell}\,\partial f_\nu/\partial x_\ell$. In dimensions $\nu < 3$,
the Lenard-Balescu equation is ultraviolet finite, and we have
\begin{eqnarray}
  \frac{\partial f_\nu}{\partial t} + 
  {\bf v}_\nu \! \cdot \!{\bm\nabla} f_\nu 
  &=& 
  L_\nu[f] ~~~~ : \nu < 3 \ .
\label{LBEsimpnu}
\end{eqnarray}
The explicit form of the scattering kernels $B_\nu[f]$ and $L_\nu[f]$
will not be required until the next lecture. As we shall see, the
scattering kernels $B_\nu$ and $L_\nu$ are simply the obvious
generalizations of their three dimensional counter parts: momentum and
wave-number vectors live in \hbox{$\nu$-dimensions}, and the
scattering is produced by the $\nu$-dimensional Coulomb potential
(\ref{Vnu}). Calculations using (\ref{BEsimpnu}) in $\nu>3$ and
(\ref{LBEsimpnu}) in $\nu<3$, respectively, are completely finite.

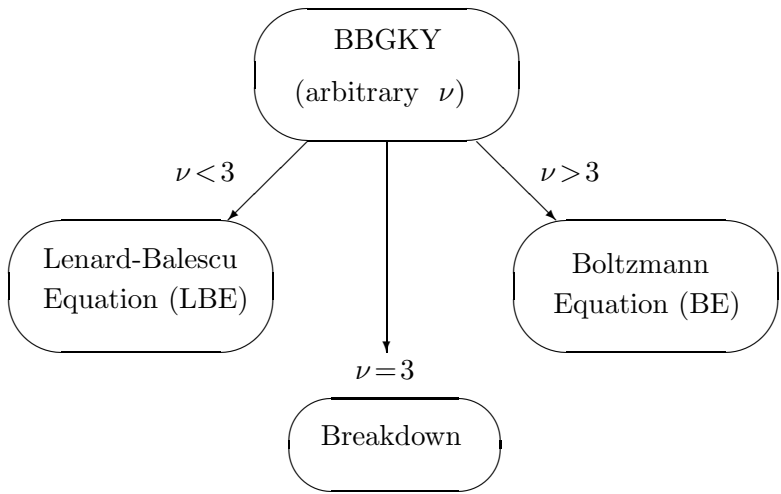
\begin{figure}[t]
\begin{picture}(120,120)(-75,0)
\put(-20,80){\oval(100,50)}
\put(-40,90){\text{BBGKY}}
\put(-55,70){(\text{arbitrary } $\nu$) }

\put(-50,55){\vector(-1,-1){30}}
\put(14,55){\vector(1,-1){30}}
\put(-100,40){$\nu\!<\!3$}\
\put(38,40){$\nu\!>\!3$}

\put(50,5){\text{Boltzmann}}
\put(43,-10){\text{Equation (BE)}}
\put(78,0){\oval(100,50)}

\put(-150,7){\text{Lenard-Balescu}}
\put(-150,-8){\text{Equation (LBE)}}
\put(-113,0){\oval(100,50)}

\put(-20,55){\vector(0,-1){80}}
\put(-32,-35){$\nu\!=\!3$}
\put(-45,-60){\text{Breakdown}}
\put(-17,-60){\oval(80,35)}

\end{picture}
\vskip3cm 
\caption{\captionskip
  For $\nu\!>\!3$ the ``textbook derivation'' of the Boltzmann
  equation for a Coulomb potential is rigorous; furthermore, the BBGKY
  hierarchy reduces to the Boltzmann equation to leading order in
  $g$. A similar reduction from the BBGKY hierarchy holds for the
  Lenard-Balescu equation in $\nu \!<\! 3$, and the ``textbook
  derivation'' is also rigorous in these dimensions. In $\nu=3$, the
  {\em derivations} of the Boltzmann and Lenard-Balescu equations
  break down for the Coulomb potential. This is not because the three
  dimensional BBGKY hierarchy breaks down, but because the Boltzmann
  and Lenard-Balescu equations break down. Indeed, BBGKY is completely
  finite for the Coulomb potential in $\nu=3$, albeit completely
  useless for our purposes.
}
\label{fig:bbgky}
\end{figure}

In exactly the same manner, one can generalize the BBGKY hierarchy to
an arbitrary number of dimensions, and this will be the starting point
for my treatment of dimensional continuation. As we discussed in the
closing paragraph of the last section, when the number of spatial
dimensions is greater than three, BBGKY reduces to the Boltzmann
equation (\ref{BEsimpnu}) to leading order in the plasma coupling
$g$. Therefore, when $\nu>3$, to leading order in $g$ the rate becomes
\begin{eqnarray}
  \frac{d{\cal E}_\nu^\smGT}{dt}
  =
  \int \!d^{\,\nu}v ~\frac{1}{2}\, m v_\nu^2 ~
  B_\nu[f] ~~~~ : ~~ \nu>3 \ .
\label{dedtgtthree}
\end{eqnarray}
Conversely, in dimensions $\nu<3$, BBGKY reduces to the Lenard-Balescu
equation (\ref{LBEsimpnu}) to leading order in $g$, and
\begin{eqnarray}
  \frac{d{\cal E}_\nu^\smLT}{dt}
  =
  \int \!d^{\,\nu}v ~\frac{1}{2}\, m v_\nu^2 ~
  L_\nu[f] ~~~~ : ~~ \nu<3 \ .
\label{dedtltthree}
\end{eqnarray}
This is illustrated in Fig.~\ref{fig:bbgky}.  The ``greater-than'' and
``less-than'' superscripts are to remind us that the integrals in the
rate calculations (\ref{dedtgtthree}) and (\ref{dedtltthree}) are to
be preformed in $\nu>3$ and $\nu<3$, respectively.  Because of spatial
uniformity, I have set the convective terms in (\ref{BEsimpnu}) and
(\ref{LBEsimpnu}) to zero, ${\bf v}_\nu \!\cdot\!{\bm\nabla} f_\nu =
0$.  Using the appropriate Coulomb potential (\ref{Vnu}) for
$V_\nu(r)$ in the scattering kernels of (\ref{dedtgtthree}) and
(\ref{dedtltthree}), the integrals now converge, and they are
calculated exactly in Sections~7 and 8 of BPS~\cite{bps}. In the
next lecture, we will calculate these integrals in the Born
approximation. 

Note that this is a first-principles derivation of the rates
(\ref{dedtgtthree}) and (\ref{dedtltthree}) in their respective
dimensions $\nu$. Let me reiterate the argument once again, although
with a slightly different emphasis. For simplicity, we consider the
purely classical regime first, adding quantum mechanics in a
moment. The classical BBGKY hierarchy for the Coulomb potential in
three spatial dimensions is well defined and finite. We run into
trouble only when we attempt to truncate the hierarchy and derive
lower-order kinetic equations, such as the Boltzmann and the
Lenard-Balescu equations. The necessity of truncated equations is of
course clear (even the three-body problem cannot be solved
analytically). Unfortunately, however, the Coulomb potential in three
dimensions produces divergent scattering kernels in these truncated
equations. Rather than creating a model of the ostensibly divergent
scattering kernel, we shall instead systematically regulate the
divergences by letting the spatial dimension depart from its
empirically measured value of three. Logarithmic divergences in three
dimensions then become simple poles of the form $1/(\nu-3)$ in
arbitrary dimensions. As with the regularization procedure of quantum
field theory, our starting point here is a {\em well-defined and
finite theory}, albeit in $\nu$ dimensions, {\em regularized in a
consistent fashion at all length and energy scales}.

Let us now return to classical BBGKY, but this time in a spatial
dimension $\nu$ of arbitrary positive {\em integer} value (we are not
yet considering continuous values of $\nu$). At first sight, the
hierarchy equations in $\nu$ dimensions are just as useless as those
in three \hbox{dimensions --- there} are simply too many of them to
solve. However, if we are willing to work to leading order accuracy in
the plasma coupling $g$, which is quite accurate for a weakly to
moderately coupled plasma, then: (i) in $\nu>3$ we can truncate BBGKY
to the $\nu$-dimensional Boltzmann equation, and (ii) in $\nu<3$ we
can truncate BBGKY to the $\nu$-dimensional Lenard-Balescu equation.
Quantum scattering effects in the plasma will not modify the
$\nu$-dimensional Lenard-Balescu equation, but they will modify the
$\nu$-dimensional Boltzmann equation. Since we require the leading
order term in the rate to be exact, we must include two-body quantum
scattering effects exactly (but no more than two-body effects, since
these are subleading \hbox{in $g$ --- in} fact, three-body and higher
correlations and scattering effects, both classical and quantum, can
and {\em should} be neglected to leading order in $g$). Two-body
quantum effects can be accounted for by replacing the classical cross
section in the Boltzmann scattering kernel with the the two-body
quantum cross section (this can be performed {\em exactly} since all
scattering phase shifts $\delta_\ell$ are known for the Coulomb
potential in three dimensions). These ideas will be illustrated in
complete detail in the next lecture for a particularly useful but
simple case.

\pagebreak
\subsection{Calculating the Leading Order Term} 
\label{sec:calcleading}

We now return specifically to the electron-ion temperature
equilibration rate of (\ref{dedteI}).  To obtain the leading order in
$g$ behavior when $\nu>3$, we calculate the rate using
(\ref{dedtgtthree}). As our calculation in the next lecture will
reveal, this rate is proportional to $g^2$ (or the number density) and
takes the form\footnote{
\footnoteskip \label{foot:gnudef}
  As we have already pointed out in footnote~\ref{foot:smallg}, the
  parameter $g=e^2 \kappa_\smD/4\pi T$ given by (\ref{gdefA}) is
  dimensionless only for $\nu=3$.  From definition (\ref{gdef}) and
  the Coulomb potential (\ref{Vnu}), the proper {\em dimensionless}
  expansion parameter should be $g_\nu= e^2 \kappa_\smD^{\nu-2}/
  C_\nu\,T$. However, since $g_\nu = g \cdot \kappa_\smD^{\nu-3}(4\pi/
  C_\nu)$, we can absorb factors of $\kappa_\smD^{\nu-3} (4\pi/C_\nu)$
  into any accompanying coefficient, such as $H(\nu;\eta)$ in
  (\ref{dedtonecal}). Therefore, powers of the dimensionless coupling
  $g_\nu$ also count powers of the three dimensional coupling
  $g$. We are therefore free to think of the expansion, even in
  $\nu$ dimensions, in terms of the three dimensional parameter $g$. 
}
\begin{eqnarray}
  \frac{d{\cal E}_{e\smI}^\smGT}{dt}
  &=& 
  H(\nu; \eta)\,\frac{g^2}{\nu-3} 
  +
  {\cal O}(\nu-3) 
  ~~:~  \text{LO in $g$ when }\nu > 3 \ ,
\label{dedtonecal}
\end{eqnarray}
where, for definiteness, we have restored the electron-ion subscript
to the rate, as in (\ref{dedtei}).  In the next lecture we shall
calculate $H(\nu; \eta)$ explicitly, but for now it will suffice to
note that $H$ depends upon the spatial dimension $\nu$ and the quantum
parameter $\eta$.  The simple pole at $\nu=3$ reflects the
long-distance or infrared divergence of the Coulomb potential in three
spatial dimensions, and it arises from an integral over the radial
coordinate of the \hbox{$\nu$-dimensional} Coulomb potential
(\ref{Vnu}).  In a similar manner, the leading order behavior in
dimensions $\nu<3$ is given by (\ref{dedtltthree}). In the next
lecture, we will see that this takes the form
\begin{eqnarray}
  \frac{d{\cal E}_{e\smI}^\smLT}{dt}
  &=&
  G(\nu)\, \frac{g^{\nu-1}}{3-\nu} 
  + {\cal O}(3-\nu) 
  ~~:~ \text{LO in $g$ when } \nu < 3 \ .
\label{dedttwocal}
\end{eqnarray}
There is no \hbox{$\eta$-dependence} in $G(\nu)$ since the leading
order long-distance physics is purely classical. In the next lecture,
we will preform the integrals to establish (\ref{dedtonecal}) and
(\ref{dedttwocal}), thereby calculating the coefficients $H(\nu)$ and
$G(\nu)$ exactly (for notational simplicity, I will hereafter drop the
\hbox{$\eta$-dependence} from $H$).  Note that (\ref{dedttwocal}) also
contains a simple pole at $\nu=3$ arising from an integration of
(\ref{Vnu}), but this time the pole corresponds to missing
short-distance physics or the ultraviolet divergence of the
Lenard-Balescu equation in three dimensions. In general, if a
three-dimensional integral diverges, then the corresponding
\hbox{$\nu$-dimensional} integral will typically contain a simple pole
of the form $1/(\nu-3)$. In this way we can transform a divergent
integral into a convergent quantity that we can manipulate, and this
is how we shall regularize the divergent Boltzmann and Lenard-Balescu
equations in three dimensions.

\subsection{Next-to-Leading Order from Leading Order via Analytic Continuation}

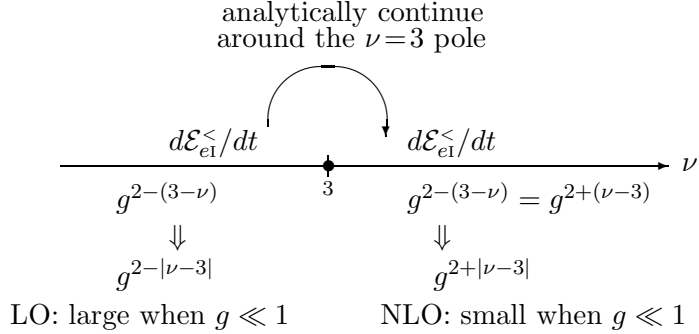
\begin{figure}[t]
\begin{picture}(120,120)(-75,0)
\put(-101,60){\vector(1,0){230}}
\put(134,58){$\nu$}

\put(0,56){\line(0,1){8}}
\put(-2.3,57.3){$\bullet$}
\put(-2.0,49){${\scriptstyle 3}$}

\put(-60,66){$d{\cal E}_{e\smI}^\smLT/dt$}
\put(-80,44){$g^{2-(3-\nu)}$}
\put(-60,30){$\Downarrow$}
\put(-80,15){$g^{2-\vert \nu-3 \vert}$}
\put(-120,0){LO: large when $g \ll 1$}

\put(30,66){$d{\cal E}_{e\smI}^\smLT/dt$}
\put(30,44){$g^{2-(3-\nu)}=g^{2+(\nu-3)}$}
\put(40,30){$\Downarrow$}
\put(40,15){$g^{2+\vert \nu-3 \vert}$}
\put(20,0){NLO: small when $g \ll 1$}

\put(0,75){\oval(45,45)[t]}
\put(22,72){\vector(0,-1){1}}
\put(-40,115){analytically continue}
\put(-42,105){around the $\nu\!=\!3$ pole}

\end{picture}
\caption{\captionskip
The analytic continuation of $d{\cal E}_{e\smI}^\smLT/dt$ from
$\nu<3$ to the region $\nu>3$: the same expression can be used
for $d{\cal E}_{e\smI}^\smLT/dt$ throughout the complex plane 
since the pole at $\nu=3$ can easily be avoided. Note that the
quantity $d{\cal E}_{e\smI}^\smLT/dt\sim g^{2+(\nu-3)}$ 
is leading order in $g$ for $\nu < 3$. However, upon analytically 
continuing to $\nu>3$ we find that $d{\cal E}_{e\smI}^\smLT/dt\sim 
g^{2+\vert \nu-3\vert}$, which is next-to-leading order in $g$  
relative to $d{\cal E}_{e\smI}^\smGT/dt \sim g^2$. 
}
\label{fig:NLOdedt}
\end{figure}

Since the rates $d{\cal E}_{e\smI}^\smGT/dt$ of (\ref{dedtonecal}) and
$d{\cal E}_{e\smI}^\smLT/dt$ of (\ref{dedttwocal}) were calculated in
mutually exclusive dimensional regimes, one might think that they
cannot be compared. However, even though (\ref{dedttwocal}) was
originally calculated in $\nu<3$ for integer values of $\nu$, we can
analytically continue the result to complex values of $\nu$ (in the
same way that the factorial function, which operates on positive
integers, can be generalized to the Gamma function, which operates on
the whole complex plane).  In fact, if we continue $d{\cal
E}_{e\smI}^\smLT/dt$ to real values of $\nu$ with $\nu>3$, then we can
directly compare (\ref{dedtonecal}) and (\ref{dedttwocal}). Upon
writing the $g$-dependence of (\ref{dedttwocal}) as $g^{2 + (\nu-3)}$,
when $\nu>3$ we see that (\ref{dedttwocal}) is higher order in $g$
than (\ref{dedtonecal}). By power counting arguments, there are no
powers of $g$ in (\ref{dedtonecal}) between $g^2$ and $g^{\nu-1}$ for
$\nu>3$, and therefore (\ref{dedttwocal}) indeed provides the correct
next-to-leading order term in $g$ when the dimension is analytically
continued to $\nu>3$,
\begin{eqnarray}
  \frac{d{\cal E}_{e\smI}^\smLT}{dt}
  &=&
  -G(\nu)\, \frac{g^{2+(\nu-3)}}{\nu-3} 
  +
  {\cal O}(\nu-3) 
  ~~:~\text{NLO in $g$ when } \nu > 3 \ ,
\label{NLOgterm}
\end{eqnarray}
where I have written the exponent of $g$ as $2+(\nu\!-\!3)$ rather
than $\nu-1$.  This is illustrated in Fig.~\ref{fig:NLOdedt}.  

As we shall see in the next section, the individual pole-terms in
(\ref{dedtonecal}) and (\ref{NLOgterm}) will cancel giving a finite
result when the leading and next-to-leading order terms are added. The
resulting finite quantity will therefore be accurate to leading and
next-to-leading order in $g$ as the $\nu \to 3$ limit is taken.
Alternately, we could have analytically continued (\ref{dedtonecal})
to dimensions $\nu < 3$, where it would become next-to-leading order
relative to (\ref{dedttwocal}). In either case, the leading and
next-to-leading order contribution can be found by simply adding
(\ref{dedtonecal}) and (\ref{dedttwocal}) and then taking $\nu \to 3$.

\subsection{Returning to Three Dimensions}
\label{sec:cartoon}

To find the three dimensional rate $d{\cal E}_{e\smI}/dt$, accurate to
leading and next-to-leading order in the plasma coupling $g$, we add
the leading order expression (\ref{dedtonecal}) for $d{\cal E}_{e
\smI}^\smGT/dt$ to the next-to-leading order expression
(\ref{NLOgterm}) for $d{\cal E}_{e\smI}^\smLT/dt$, and then take the
limit $\nu \to 3^+$:
\begin{eqnarray}
  \frac{d{\cal E}_{e\smI}}{dt}
  =
  \lim_{\nu \to 3^+}
  \left[
  \frac{d{\cal E}_{e\smI}^\smGT}{dt}
  +
  \frac{d{\cal E}_{e\smI}^\smLT}{dt}
  \right] + {\cal O}(g^3) \ .
\label{dedtorderg3}
\end{eqnarray}
For the same reasons as given in Section~\ref{sec:assembly}, this does
not lead to any form of ``double counting.''  Instead, we are simply
adding the next-to-leading order term (\ref{NLOgterm}) to the leading
order term (\ref{dedtonecal}) at a common value of~\hbox{$ \nu > 3$}.

Since (\ref{dedtorderg3}) lies at the heart of dimensional
continuation, allow me review the reasoning behind this expression one
final time.  Recall from (\ref{dedtonecal}) and (\ref{dedttwocal})
that $d{\cal E}_{e\smI}^\smGT/dt$ and $d{\cal E}_{e\smI}^\smLT/dt$ are
both leading order in $g$ for $\nu>3$ and $\nu<3$, respectively. Since
these functions were calculated for mutually exclusive values of
$\nu$, they must be analytically continued to the same value of $\nu$
for purposes of comparison. In equation (\ref{dedtorderg3}), we have
chosen to continue $d{\cal E}_{e\smI}^\smLT/dt$ to $\nu>3$, which
takes the same algebraic form as it did for $\nu<3$. The analytic
continuation has the effect of rendering $d{\cal E}_{e\smI}^\smLT/dt$
subleading in $g$ relative to $d{\cal E}_{e\smI}^\smGT/dt$. There are
no powers of $g$ between these two terms for any value of $\nu$, and
therefore the limiting procedure $\nu \to 3^+$ gives the three
dimensional rate {\em exactly} to leading and next-to-leading order
accuracy in $g$.  We have therefore found the leading order and
next-to-leading order in $g$ contributions for $\nu>3$,
\begin{eqnarray}
  \frac{d{\cal E}_{e\smI}^\smGT}{dt}
  +
  \frac{d{\cal E}_{e\smI}^\smLT}{dt}
  =
  \frac{1}{\nu-3} \Big[H(\nu)\, g^2 - G(\nu)\, g^{2+(\nu-3)}\Big]
  +
  {\cal O}(\nu-3) \ .
\label{NLOsum}
\end{eqnarray}
I have not indicated the higher order error in $g$, but instead I am
emphasizing here the error associated with the $\nu$-expansion. That
is to say, (\ref{NLOsum}) contains an implicit error that approaches
${\cal O}(g^3)$ as $\nu \to 3^+$, while I have chosen to explicitly
display the first-order error in $\nu-3$.  In the sum (\ref{NLOsum}),
the error in $g$ remains nonzero in the three dimensional limit, while
the $\nu$-error in (\ref{NLOsum}) vanishes as $\nu \to 3$.

In taking the $\nu \to 3$ limit, we must keep in mind that terms
proportional to $\nu\!-\!3$ will give a non-vanishing result when
multiplied by the pole $1/(\nu-3)$. Upon expanding the
\hbox{$g$-dependence} to linear order in $\epsilon=\nu\!-\!3$ we find
\begin{eqnarray}
  g^{\epsilon} 
  = 
  \exp\{\,\ln\!\left(g^\epsilon\right)\}
  = 
  \exp\{\epsilon \ln g\} 
  = 
  1 + \epsilon\ln g + {\cal O}(\epsilon)^2 \ ,
\end{eqnarray}
which can be written as
\begin{eqnarray}
  \frac{g^\epsilon}{\epsilon}
  = 
  \frac{1}{\epsilon} 
  +
  \ln g + {\cal O}(\epsilon) \ .
\label{gexp}
\end{eqnarray}
We must now expand the coefficients $H(\nu)$ and $G(\nu)$ in powers of
$\epsilon=\nu-3$. As we shall see in the next lecture, and this is a
{\em crucial} point, the leading order in $\epsilon$ terms are
identical, and so the expansions take the form
\begin{eqnarray}
  H(\nu) 
  &=&
  -A + \epsilon \,H_1 + {\cal O}(\epsilon^2)
\label{Hexp}
\\[5pt]
  G(\nu) 
  &=&
  -A + \epsilon \,G_1 + {\cal O}(\epsilon^2) \ .
\label{Gexp}
\end{eqnarray}
In the next lecture, we will perform the integrals in
(\ref{dedtgtthree}) and (\ref{dedtltthree}), thereby allowing us to
exactly compute $A \equiv H(\nu\!=\!3) = G(\nu\!=\!3)$ and the linear
terms $H_1 \equiv H^\prime(\nu\!=\!3)$ and $G_1 \equiv
G^\prime(\nu\!=\!3)$. The remaining procedure is now straightforward.
Upon using (\ref{gexp})--(\ref{Gexp}) in (\ref{dedtonecal}) and
(\ref{NLOgterm}), the leading and next-to-leading order terms in $g$
now take the form
\begin{eqnarray}
  \frac{d{\cal E}_{e\smI}^\smGT}{dt}
  &=& 
  -
  \frac{A}{\nu-3} \, g^2
  +
  H_1\,g^2
  +
  {\cal O}(\nu-3;g^3)
\label{LOcalcA}
\\[5pt]
  \frac{d{\cal E}_{e\smI}^\smLT}{dt}
  &=& 
    \phantom{-}
  \frac{A}{\nu-3} \, g^2 
  -
  A\, g^2 \ln g - G_1 \,g^2
  +
  {\cal O}(\nu-3;g^3) \ .
\label{LOcalcB}
\end{eqnarray}
I have now indicated here the error associated with the $g$-expansion.
Because the leading order terms in (\ref{Hexp}) and (\ref{Gexp}) are
equal, the simple poles in (\ref{LOcalcA}) and (\ref{LOcalcB})
cancel. Therefore, the limit $\nu \to 3^+$ of expression
(\ref{NLOsum}) gives the rate
\begin{eqnarray}
  \frac{d{\cal E}_{e\smI}}{dt}
  =
  - A\, g^2 \ln g   +  B\, g^2  + {\cal O}(g^3) \ ,
\end{eqnarray}
with $B=H_1-G_2$, in agreement with (\ref{dedtNLO}). In this way, 
BPS has calculated the energy exchange accurate to leading order 
and next-to-leading order in $g$.

\section{Some Closing Remarks}

\subsection{Summary}

I hope this account of the simple field theoretic apparatus necessary
for a reading of BPS has been helpful. Having understood the reasoning
behind BPS, especially the manner by which analytic continuation turns
a leading order result into a next-to-leading order result, we can
summarize the BPS procedure by the following prescription:

\begin{enumerate}
\enumerateskip

\item[1.] Calculate the rate $d{\cal E}^\smGT\!/dt$ in the regime
  $\nu>3$ using the Boltzmann equation, including quantum corrections
  to all orders in $\eta$. This captures the leading order physics in
  dimensions greater than three.

\item[2.] Calculate the corresponding rate $d{\cal E}^\smLT\!/dt$ in
  the regime $\nu<3$ using the Lenard-Balescu equation. This physics
  is classical, and captures the leading order behavior in dimensions
  less than three.

\item[3.] Add the terms $d{\cal E}^\smGT\!/dt$ and $d{\cal E}^\smLT\!/dt$ 
  and take the $\nu \to 3$ limit. This gives the energy exchange
  rate in three dimensions accurate to leading and next-to-leading 
  order in $g$, $d{\cal E}/dt =-A g^2 \ln\{C(\eta) g \} + 
  {\cal O}(g^3)$. That is to say, this gives the coefficients  $A$ 
  and $C$ exactly. 
  
\end{enumerate}

\noindent
This prescription provides complete analytic expressions for the
coefficients $A$ and $C(\eta)$, with $C(\eta)$ being of particular
interest.  In the language of the Coulomb logarithm, we write $\ln
\Lambda_{\smCoul}(g,\eta)=-\ln\{C(\eta) g\}$.

\subsection{Further Context}

The fundamental interactions of nature can be expressed as quantum
field theories (except for gravity\footnote{\footnoteskip
  There have of course been numerous attempts at quantizing gravity,
  but since none of these has yet produced a full fledged theory of
  quantum gravity, I do not count them here. In fact, it might not be
  possible to describe gravity with a field theory on the quantum
  level. This would be true, for example, if string theory were to
  provide the theory of quantum gravity, since a string theory is
  qualitatively different from a field theory.
}, 
which to date has only been expressed as a classical field theory,
{\em i.e.}  general relativity). For this reason, quantum field theory
reveals something very deep about the structure of nature. However,
for our purposes, we only need to think of quantum field theory as an
elegant collection of tools for packaging and solving many-body
problems, and plasma physics concerns itself with the many-body
problem par excellence. It should therefore come as little surprise
that quantum field theory can be useful for plasma physics.
Reference~\cite{bps} is a nice example of cross fertilization between
two quite different branches of physics, and it is gratifying for
someone who has worked in both particle and plasma physics that such
seemingly different subjects can inform one another. Perhaps the most
dramatic difference between particle and plasma physics lies with
their respective methods and outlooks, and not so much the subject
matter itself. A number of plasma physicist have explained to me that
that plasma physics is not an {\em exact} science. While the nature of
the subject might render a first-principles approach limited,
particularly for subtopics like magneto-hydrodynamics in tokamaks or
self-organized behavior in strongly coupled plasmas, I believe the
plasma physicist could still benefit from the more rigorous outlook of
particle physics. Conversely, the particle physicist could benefit
from plasma physics. In fact, until new high energy experiments come
on-line, plasma physics may have more to offer particle physics than
the reverse.\footnote{\footnoteskip
  Particle physics is currently in crisis brought on from a dearth,
  and indeed a complete absence, of new experimental results. The only
  hope for high energy physics seems to lie with the Large Hadron
  Collider at CERN, or its possible successors. Compare this situation
  with the current Renaissance in observational cosmology and
  astrophysics, which has more and better quality data than at any
  time during its history. Sadly,~the field of high energy physics has
  responded to its data crisis in a rather pathological, although
  perhaps predictable, manner: the mono-culture of string theory and
  string theory ``inspired'' theories, none of which are empirically
  based, and all of which virtually dominate the entire landscape of
  high energy physics. At the 10\% level, string theory and the like
  might be healthy (after all, a quantum description of gravity simply
  falls out of string theory, and it is hard to believe that this is
  an accident); however, at the current level, string theory and its
  kinfolk are as stifling as a hillside of kudzu in Tennessee.
}
Until recently, high energy physics was a spectacular field in 
which to work, but, with a few exceptions such as lattice gauge 
theory and RHIC physics at BNL, the field seems to have been milked 
dry. High energy physics is in dire need of direct experimental 
input. Plasma physics, on the other hand, still has many interesting 
and quite challenging problems of experimental and astrophysical 
relevance, and a number of these seem tailor-made for the field 
theorist. The particle physicist could therefore benefit by an 
excursion (or even a longer stay) in plasma physics. Hence, my 
hidden agenda in these lectures: it would be quite nice to entice 
a few particle theorists to work on some of the interesting problems 
of plasma physics. It would also be nice to convince a few plasma 
physicists that particle theory has something to offer them as well.

\begin{acknowledgments}
  This manuscript arose from a series of lectures presented to the
  Applied Physics Division's Plasma Working Group in August 2006 at
  the Los Alamos National Laboratory. I would kindly like to thank Guy
  Dimonte for providing the opportunity to speak, and the other
  members of the Plasma Working Group for their questions and interest
  during the talks. I would particularly like to thank Jerome
  Daligault for reading the manuscript for clarity and accuracy,
  and Lowell Brown for a number of useful discussions.  This work was
  funded by the ICF Codes Projects JA1K and JL1C, and the
  Thermonuclear Burn Project JAVD.
\end{acknowledgments}




\end{document}